\documentclass{article}

\usepackage{listings}

\lstdefinestyle{matlabstyle}{
    language=Matlab,
    basicstyle=\ttfamily\footnotesize,
    keywordstyle=\color{blue},
    commentstyle=\color{green!50!black},
    stringstyle=\color{red},
    numbers=left,
    numberstyle=\tiny\color{gray},
    stepnumber=1,
    breaklines=true,
    captionpos=b,
    morekeywords={factorial, nchoosek, setdiff}
}

\usepackage{mathrsfs}
\usepackage{lmodern}
\usepackage[table,xcdraw]{xcolor}
\usepackage{colortbl}
\usepackage{graphicx}
\usepackage{multirow}%
\usepackage{amsmath,amssymb,amsfonts}%
\usepackage{amsthm}%
\usepackage[title]{appendix}%
\usepackage{float}
\usepackage{textcomp}%
\usepackage{manyfoot}%
\usepackage{booktabs}%
\usepackage{algorithm}%
\usepackage{algorithmicx}%
\usepackage{algpseudocode}%
\usepackage{listings}%
\usepackage{soul}
\usepackage{hyperref}
\usepackage{xurl}
\usepackage{caption}
\title{\Large \textbf {Principal Component Analysis When \textit{n} $<$ \textit{p}: Challenges and Solutions \\ \text{---------- PRE - PRINT VERSION ----------}}}

\makeatletter
\renewcommand\@date{{%
  \vspace{-\baselineskip}%
  \large\centering

  \begin{tabular}[t]{c@{\extracolsep{1em}}c} 
    \textbf{Nuwan Weeraratne} \\
      \normalsize Dept. of Mathematics and Statistics \\
      \normalsize University of Waikato, New Zealand \\
    \normalsize ncweera.nzinfo@gmail.com
  \end{tabular}%
    \begin{tabular}[t]{c@{\extracolsep{1em}}c} 
    \textbf{Lyn Hunt} \\
      \normalsize Dept. of Mathematics and Statistics \\
      \normalsize University of Waikato, New Zealand  \\
    \normalsize lah@waikato.ac.nz
  \end{tabular}

   \bigskip
  \begin{tabular}[t]{c@{\extracolsep{2em}}c} 
    \textbf{Jason Kurz} \\
      \normalsize Dept. of Mathematics and Statistics \\
      \normalsize University of Waikato, New Zealand \\
    \normalsize jason.kurz@waikato.ac.nz
  \end{tabular}

}}
\makeatother

\begin{document}

\maketitle

\begin{center}
\textbf{Abstract} \\
\end{center}

Principal Component Analysis is a key technique for reducing the complexity of high-dimensional data while preserving its fundamental data structure, ensuring models remain stable and interpretable. This is achieved by transforming the original variables into a new set of uncorrelated variables (principal components) based on the covariance structure of the original variables. However, since the traditional maximum likelihood covariance estimator does not accurately converge to the true covariance matrix, the standard principal component analysis performs poorly as a dimensionality reduction technique in high-dimensional scenarios $n<p$. In this study, inspired by a fundamental issue associated with mean estimation when $n<p$, we proposed a novel estimation called pairwise differences covariance estimation with four regularized versions of it to address the issues with the principal component analysis when \textit{n} $<$ \textit{p} high dimensional data settings. In empirical comparisons with existing methods (maximum likelihood estimation and its best alternative method called Ledoit-Wolf estimation) and the proposed method(s), all the proposed regularized versions of pairwise differences covariance estimation perform well compared to those well-known estimators in estimating the covariance and principal components while minimizing the PCs' overdispersion and cosine similarity error. Real data applications are presented.\\


\section{Introduction}\label{sec1}

During the last two decades, technology has developed rapidly, minimizing the processing time of most statistical applications and algorithms. Because technology advances rapidly, researchers constantly introduce new technological concepts and methodologies into the research world. In that context, Machine Learning (ML) is a trending topic in the research and development (R\&D) industry. As Arthur Samuel said in 1959, ML is a vital part of Artificial Intelligence (AI) that focuses on developing a statistical algorithm that can learn from data and generalize to new unsighted data without detailed human guidance or instruction \cite{samuel1959some}. Supervised, unsupervised, semi-supervised and reinforcement learning are the four main ML methods used in practice. A wide range of fields use ML algorithms. However, the choice of an ML algorithm depends on the characteristics of the data, expected usage, desired performance, domain of the statistical application, and other factors.\\

However, ML algorithms are now often used to operate high-dimensional data in various applications in diverse fields, such as bioinformatics, forensic science, cyber security, agriculture, healthcare, finance, and computer science. However, fitting models in high-dimensional spaces requires substantial resources and time. The challenge Strengthens due to the "Curse of Dimensionality," \cite{Bellman+1961} where increasing dimensions lead to sparsity in data, complicating model training, and potentially causing overfitting. Regularization or dimensional reduction (DR) techniques are essential to mitigate this challenge. DR helps compress data while retaining essential information, facilitating more efficient and effective model-building processes across domains.\\

DR is a fundamental technique used in intermediate data analysis and Unsupervised ML algorithms, transforming the original data into a lower-dimensional space while preserving its essential characteristics. These algorithms find hidden patterns in the given data without any detailed human guidance or instruction. To gain better insights into high-dimensional data, it enhances the effectiveness of ML models by simplifying the structure of data, eliminating redundancy, reducing computation time, and improving visualization capability. In general, DR techniques can be divided into two main types: feature selection and feature extraction. The feature selection method comprises selecting the most critical subset of the original dimensions from the existing dataset \cite{guyon2003introduction}. It does not create any new dimensions based on the existing dataset. Meanwhile, the Feature extraction method creates new dimensions that are linear combinations of the original dimensions by transforming the original high-dimensional data into a lower-dimensional space \cite{jolliffe2002principal}. Researchers use feature extraction methods more widely than feature selection methods. Several feature extraction methods exist, each with its unique approach and underlying principles. Some of the most widely used feature extraction techniques are principal component analysis (PCA), linear discriminant analysis, t-distributed stochastic neighbor embedding (t-SNE), and etc.\\

PCA is the widely used method for DR among these feature extraction methods. PCA ensures reasonable data representation by transforming original high-dimensional data into a lower-dimensional space while keeping most of the total variance. When dealing with high-dimensional data, PCA facilitates more efficient ML algorithms by meaningfully shaping the data. Otherwise, the data's redundancy minimizes the effectiveness of most high-dimensional statistical approaches.\\

In recent years, there has been significant growth in utilizing high-dimensional data in situations where fewer observations than the number of dimensions. However, existing most commonly used DR techniques, including PCA, have not been effective in such kind of situations because most of the DR methods rely on the mean estimation, and with limited observations, the mean is more likely to be influenced by random noise and outliers, leading to inaccurate centering. This challenge has served a significant research interest to the researchers who focused mostly on issues of high-dimensional data analysis. Therefore, this research seeks to deeply study the effectiveness of PCA in these scenarios and understand how it can be optimized to handle the challenges posed by \textit{n} $<$ \textit{p} high-dimensional data settings. By doing so, the goal is to enhance our ability to extract meaningful insights from high-dimensional data despite the limitations imposed by the insufficient sample size and the number of dimensions.

\section{Problem Statement}\label{sec2}

PCA transforms the original dataset into a differently structured dataset by using linear combinations of original dimensions while keeping the original dataset's variance-covariance structure as much as possible. Therefore, PCA depends on the variance-covariance structure of the dataset and uses the Maximum Likelihood Estimation (MLE) of covariance estimation to obtain the sample variance-covariance structure of the original data set. Although traditional covariance MLE is recognized for its asymptotic unbiasedness, its efficacy becomes notably challenged in high-dimensional scenarios where \textit{n} $<$ \textit{p}. This problem arises because MLE needs more observations (\textit{n}) than the number of dimensions (\textit{p}). When in \textit{n} $<$ \textit{p} high-dimensional situations, the mean is more likely to be influenced by random noise and outliers, leading to inaccurate centering, and this increased variability translates into less reliable estimates of the actual central tendency of each dimension. That means, in \textit{n} $<$ \textit{p} high-dimensional data settings, when \textit{n} goes down compared to \textit{p}, the Mean Squared Error (MSE) of the sample covariance matrix gets higher. This kind of higher MSE implies that the sample covariance matrix is not an accurate and reliable estimator for the true covariance \cite{johnstone2001distribution}, \cite{bickel2008covariance}. These inaccurate sample estimates of variance-covariance structure result in misleading sample principal component estimates in high-dimensional data where \textit{n} $<$ \textit{p}. we summarized those challenging issues concerning PCA  as follows.\\

\begin{enumerate}

\item High Cosine Similarity Error (CSE) of sample PC due to overfitting, leading to sample PCs misaligning with population PCs.

\item Mean bias of eigenvalues increases as 
\textit{n} decreases, causing inaccurate estimation of PCs.

\item Overestimated first \textit{n}-1 eigenvalues and underestimated last 
\textit{p}-\textit{n}+1 eigenvalues, leading to misinterpretation.

\item Rank deficiency in sample covariance matrix limits the number of independent PCs to at most \textit{n}-1, reducing explanatory power.

\item High condition number (ratio of largest to smallest eigenvalue), making the sample covariance matrix ill-conditioned.

\item Overdispersion of explained variance in the first \textit{n}-1 PCs, capturing noise rather than true variance.

\end{enumerate}

\section{State of the Art}

\subsection{Recent Developments}
As a result, due to these issues in \pmb{$\hat{\Sigma}$}, the need of a well conditioned covariance matrix came into account. A well-conditioned estimator can reduce estimation error (projection cost) by improving the estimation of eigenvalues. Therefore, in this section, we provide a selected brief overview of the existing literature on covariance estimation techniques for high-dimensional datasets to further highlight the issue of PCA when \textit{n} $<$ \textit{p} high-dimensional data settings as follows. \\

\begin{itemize}

    \item In 1975, Charles Stein proposed a new covariance estimation method called the ``Stein Estimator" \cite{stein1975estimation} under which Stein's loss function shrinks eigenvalues towards a central value to improve estimation. It shrinks eigenvalues but lacks positive definiteness, is non-sparse, and is unstable for \textit{n} $<$ \textit{p}.\\

    \item In 2004, Ledoit and Wolf proposed a more general form of estimator, which would be a well-known stenian-type shrinkage (minimizing the MSE criterion) estimator for the \pmb{$\hat{\Sigma}$} \cite{ledoit2004well}. The Ledoit-Wolf covariance estimation uses a linear shrinkage approach to stabilize the covariance matrix. However, it uses uniform shrinkage, fails to capture the true covariance structure, and lacks sparsity.\\

    \item In 2008,  Friedman \cite{friedman2008sparse} proposed a covariance estimation method to other estimator and distribution types to extend the Dempster covariance selection problem for a multivariate Gaussian distribution when observations are constrained. It is called Graphical Lasso (GLasso) estimation and applies \textit{l}\textsubscript{1} regularization to enforce sparsity in the inverse covariance matrix. The GLasso promotes sparsity but struggles with multicollinearity and requires careful tuning of regularization.\\

    \item In 2008, Bickel and Levina developed a method of sparse estimation for the application of thresholding to off-diagonal components of a sample covariance matrix of \textit{p} dimensions estimated from \textit{n} observations \cite{bickel2008covariance}. They demonstrate the robustness of the threshold estimate in the operator norm, provided that the actual covariance matrix is sufficiently sparse, the parameters are Gaussian (or sub-Gaussian), and log \textit{p}/\textit{n} is set to zero, and explicit rates are obtained. This estimator forces sparsity but is sensitive to threshold choice and performs poorly with outliers.\\

    \item In 2008, Friedman proposed a method to estimate a sparse inverse covariance matrix using an \textit{l}\textsubscript{1} penalty. However, this method does not work well in high-dimensional data settings as a covariance estimation. The major drawback of this method is the sensitivity of the choice of penalty parameter. Improper selection of penalties leads to inaccurate and unreliable estimates of covariance. Sometimes, this \textit{l}\textsubscript{1} penalty may not fully capture the complex variance-covariance structures of the data \cite{friedman2008sparse}.\\

    \item In 2008, Levina introduced another sparse covariance estimation method for large covariance matrices using a nested lasso penalty. This estimation uses multiple layers of Lasso penalties for structured sparsity. This improves interpretability, but fails in high-dimensional settings due to sensitivity to penalty parameters \cite{levina2008sparse}.\\

    \item In 2010, Chen developed a formula that, assuming the data were Gaussian-distributed, could be used to select a decrease in the shrinkage coefficient, resulting in an MSE lower than the one reported in the 2014 Ledoit-Wolf formula \cite{chen2010shrinkage}. They improved the Ledoit-Wolf method by conditioning it with the Rao-Blackwell Ledoit-Wolf (RBLW) covariance estimation. Because RBLW Covariance Estimation uses Rao-Blackwellization to reduce bias, this approach may introduce additional computational complexity, especially for large datasets with high dimensionality. They proposed an alternative iterative approach to approximate the clairvoyant shrinkage estimator to reduce the estimation error and computational complexity further. This new method is called Oracle Approximating Shrinkage (OAS) covariance estimation \cite{chen2010shrinkage}. However, OAS reduces shrinkage bias but assumes a spherical structure and is unreliable with outliers.\\

    \item In 2011, Bien and Tibshirani suggested a Penalized Maximum Likelihood covariance estimation with a weighted lasso-type penalty method based on a sample of vectors drawn from a multivariate Gaussian distribution \cite{bien2011sparse}. However, it is sensitive to parameter choice and struggles with outliers.\\

    \item Cai and Liu introduced an adaptive variant of hard thresholding covariance estimation in 2011, originally suggested by Bickel and Levina in 2008 \cite{cai2011adaptive}. The Adaptive Thresholding Covariance Estimation is a method used to estimate covariance matrices in signal processing and machine learning. It involves adapting thresholding techniques to estimate the covariance matrix. One common approach to adaptive thresholding is to use a shrinkage estimator, such as the Ledoit-Wolf shrinkage estimator. But fails in non-sparse covariance structures.\\

    \item Xue \cite{xue2012positive} proposed a Positive Definite \textit{l}\textsubscript{1} Penalized Estimation of Large Covariance Matrices. Positive Definite \textit{l}\textsubscript{1} Penalized Estimation of Large Covariance (PDGLasso) is a method aimed at estimating high-dimensional covariance matrices that are both positive definite and sparse, which is particularly relevant when \textit{p} $>$ \textit{n}. To ensure the positive definiteness of the sample covariance matrix, PDSLasso introduces \textit{l}\textsubscript{1} penalty to the off-diagonal elements of the covariance matrix. However, this is computationally expensive and requires prior knowledge of the data structure.\\

    \item Using Convex optimization, Rothman proposed a covariance estimator called Positive Definite Sparse Covariance Estimator (PDSCE) in 2012 \cite{rothman2012positive} to generate a sparse estimate of a covariance matrix, which is positive definite, and is suitable for high-dimensional environments. However, the PDSCE struggles with rank deficiency and non-sparse structures.\\

    \item In 2013, Won \cite{won2013condition} proposed that the estimator should be maximized with the normality of the data, but the condition number constraint should be applied to the estimator. It controls the ratio of largest to smallest eigenvalues to improve conditioning. The major drawback of this covariance estimation method is its heavy reliance on condition number regularization. Therefore, it may not optimally capture the full covariance structure.\\

    \item In 2013, Fan et al. proposed a non-parametric estimator for $\pmb{\Sigma}$ based on a PCA called Principal Orthogonal ComplEment Thresholding Estimation \cite{fan2013large}. This is less effective for datasets with complex or highly dimensional covariance structures, as it assumes an approximately diagonal or nearly sparse covariance matrix, making it suboptimal for capturing strong off-diagonal dependencies.\\

    \item In 2014, Abadir et al. proposed to split data into subsets and average their covariance estimates. This improves conditioning but fails in time-varying covariance structures.\\

    \item In 2015, Ledoit and Wolf proposed a nonlinear shrinkage eigenvalue estimator for population covariance matrices that satisfies a mean-squared criterion for large-scale asymptotic functions \cite{ledoit2015spectrum}. However, it optimizes shrinkage but is complex and may not generalize to diverse covariance structures.\\

    \item In 2016, Maurya proposed a method for estimating well-defined and sparse covariance and inverse covariance from an high-dimensional sample of vectors derived from a sub-Gaussian distribution \cite{maurya2016well}. The estimators proposed are derived from the minimization of the quadratic loss function, the joint penalty of the \textit{l}\textsubscript{1} norm, and the variance of the eigenvalues of that norm [Joint PENalty Estimation of Covariance (JPEN)]. The JPEN jointly penalizes covariance and inverse covariance but relies heavily on tuning parameters.\\

    \item The Nonparametric Eigenvalue Regularized Covariance Matrix Estimation was proposed by Lam in 2016 using the same data partitioning concept in the Non-Linear Shrinkage Estimation of Large Dimensional Covariance Estimation \cite{lam2016nonparametric}. But it assumes normality and lacks robustness to outliers.\\

    \item Vu and Lei proposed two ideas to capture the sparsity of a covariance matrix. These methods facilitate the estimate of the subspace formed by the principal eigenvectors of a covariance matrix and the understanding of the variance-covariance structure of high-dimensional data. However, this method may not perform well in different types of covariance structures and \textit{n} $<$ \textit{p} high-dimensional data settings \cite{vu2013minimax}.\\

    \item In 2018, NOVEL Integration of the Sample and Thresholded Covariance Estimators was introduced by Huang and Fryzlwicz \cite{huang2019novelist}, which is a combination of linear shrinkage with sparse estimators. However, it is sensitive to the choice of the threshold parameter and the potential for bias in estimating the covariance matrix.\\

    \item In 2021, Fan and Liao proposed Sparse Factor Covariance Estimation which uses latent factor models to estimate covariance in a low-rank structure. However, it assumes sparsity, which may not hold in real data.

\end{itemize}

\subsection{Gaps in Current Understanding}

The typical MLE of covariance does not capture well the underlying variance-covariance structure when \textit{n} $<$ \textit{p} high-dimensional data analysis. However, since the 1950s, many researchers have proposed multiple alternative techniques, such as Stein’s covariance estimation, Ledoit-Wolf covariance estimation, GLasso, etc., to tackle this problem. However, all of these methods have some limitations.  Sensitivity to the shrinkage parameter, lack of sparsity, potential for non-positive definite covariance matrices, and sensitivity to outliers are common in most alternative approaches. Among them, Ledoit Wolf's estimator is more prevalent in practice. However, the primary drawback of Ledoit-Wolf covariance estimation is its uniform shrinkage approach, which may not capture the true variance-covariance structure, especially in non-sparse data. In addition, covariance estimation is the principal statistical tool we have used in PCA. As the typical MLE of covariance does not capture well the underlying variance-covariance structure, PCA also performs not well and produces misinterpretations when \textit{n} $<$ \textit{p} high-dimensional data analysis. The exciting thing is that few alternative high-dimensional covariance estimations focus on PCA as an application of covariance estimation. Therefore, there is a significant research gap in the existing literature regarding PCA when \textit{n} $<$ \textit{p} high-dimensional data settings. This gap splashes the need for a well-defined covariance estimation for PCA when \textit{n} $<$ \textit{p} high-dimensional data settings.

\section {Our Approach}

In this paper, we propose a novel sample covariance estimation method called "Pairwise Differences Covariance (PDC)", based on the idea that improving the finite sample estimates of principal components by improving the covariance estimation through increasing the sample size by taking all the pairwise differences in the sample observations, instead of sample size increasing.\\

The following algorithm estimates the novel Pairwise Differences Covariance matrix using pairwise difference computations.\\

\begin{lstlisting}[style=matlabstyle, caption={PDCEst Function}, frame=none]
function CovEst = PDCEst(data)

% Step 1: Get the dimensions of the data matrix
[n, p] = size(data);

% Step 2: Initialize PairDiff
PairDiff = zeros(n*(n-1), p);

% Step 3: Calculate pairwise differences for PairDiff
row_idx = 1;
for i = 1:n
    for j = 1:n
        if i ~= j
            PairDiff(row_idx, :) = data(i, :) - data(j, :);
            row_idx = row_idx + 1;
        end
    end
end

% Step 4: Initialize DiffData_1 and DiffData_2
DiffData_1 = [];
DiffData_2 = [];

% Step 5: Calculate the number of pairs of pairwise differences
num_pairs = factorial(n) / factorial(n - 2) + nchoosek(n - 1, 2) * n;

% Step 6: Calculate pairwise differences for DiffData_1 and DiffData_2
for i = 1:n
    pairwise_diffs = data(i, :) - data(setdiff(1:n, i), :);
    num_diffs = size(pairwise_diffs, 1);
    indices = [];
    for j = 1:num_diffs
        indices = [indices; j, j];
    end
    for j = 1:num_diffs
        for k = j+1:num_diffs
            indices = [indices; j, k];
        end
    end
    for idx = 1:size(indices, 1)
        j = indices(idx, 1);
        k = indices(idx, 2);
        if j <= num_diffs && k <= num_diffs
            DiffData_1 = [DiffData_1; pairwise_diffs(j, :)];
            DiffData_2 = [DiffData_2; pairwise_diffs(k, :)];
        end
    end
end

% Step 7: Compute DiffData
DiffData = DiffData_1' * DiffData_2;

% Step 8: Convert DiffData into a symmetric matrix
SymDiffData = 0.5 * (DiffData + DiffData');

% Step 9: Compute the covariance matrix
CovEst = (1 / (n * num_pairs)) * SymDiffData;

end
\end{lstlisting}

In general, if the data matrix has \textit{n} observations, we can have \(
\frac{n!}{(n-2)!2!}\) unique pairwise differences/combinations, where the order of selection does not matter. However, in Step 3, we compute the initial pairwise differences using a permutation technique, yielding \(n(n-1)\) pairwise differences. This results in twice the number of unique pairwise differences. In Step 6, we convert this into a matrix with \(
\frac{n!}{(n-2)!} + \binom{n-1}{2} n\) pairwise differences, multiplying the original unique pairwise differences by \textit{n} times. Therefore, we divide the final covariance by \textit{n} to obtain the correct variance estimate, ensuring we do not overestimate by a factor of \textit{n}.\\

Thus, the final covariance matrix estimate $\pmb{\hat{\Sigma}}$\textsubscript{PDC} is:

\begin{equation}
  \pmb{\hat{\Sigma}}\textsubscript{PDC} = \frac{2}{n^2(n-1)} \mathbf{E}   
\end{equation}\\


where;\\
\[
\mathbf{d}_{ij} = \mathbf{x}_i - \mathbf{x}_j
\]

\[
\mathbf{E} = \frac{1}{2} (\mathbf{M} + \mathbf{M}^T)
\]

\[
\mathbf{M} = (\mathbf{D}^{(1)})^T \mathbf{D}^{(2)}
\]

\[
\mathbf{D}^{(1)} = \begin{pmatrix}
\mathbf{D}_1^{(1)} \\
\mathbf{D}_2^{(1)} \\
\mathbf{D}_n^{(1)}
\end{pmatrix}
\quad \text{and} \quad
\mathbf{D}^{(2)} = \begin{pmatrix}
\mathbf{D}_1^{(2)} \\
\mathbf{D}_2^{(2)} \\
\mathbf{D}_n^{(2)}
\end{pmatrix}
\]

\[
\mathbf{D}_i^{(1)} = \begin{pmatrix}
\mathbf{d}_{i1} \\
\mathbf{d}_{i1} \\
\vdots \\
\mathbf{d}_{i2} \\
\mathbf{d}_{i2} \\
\vdots \\
\mathbf{d}_{i(n-1)}
\end{pmatrix}
\quad \text{and} \quad
\mathbf{D}_i^{(2)} = \begin{pmatrix}
\mathbf{d}_{i1} \\
\mathbf{d}_{i2} \\
\vdots \\
\mathbf{d}_{i2} \\
\mathbf{d}_{i3} \\
\vdots \\
\mathbf{d}_{i(n-1)}
\end{pmatrix}
\]\\

PDC was developed to improve the finite sample estimates of PCs by improving the covariance estimation by increasing the sample size by taking all the pairwise differences in the sample observations instead of the sample size. Incorporating asymptotic differences and constructing matrices \(\mathbf{D}_i^{(1)}\) and \(\mathbf{D}_i^{(2)}\) further enhances the robustness of the covariance estimation process.\\

Apart from PDC, we suggested four different regularization techniques to enhance the PDC estimation method to improve the reliability and accuracy of the estimation process where \textit{n} $<$ \textit{p} high-dimensional data settings. The four different regularized PDC estimations are [1] Standardized Pairwise Differences Covariance (SPDC), [2] Local Scaled Pairwise Differences Covariance (LSPDC), [3] Scaled by Maximum Absolute Value Pairwise Differences Covariance (MAXPDC), and [4] Scaled by Range Pairwise Differences Covariance (RPDC).\\

The SPDC approach's objective is to normalize each pairwise difference's contributions across the dataset by ensuring they have zero mean and unit variance. The regularization technique involves standardizing each pairwise difference using its mean and standard deviation.

\[
\text{Standardized\_}\mathbf{D}_{ij} = \frac{\mathbf{d}_{ij} - \mu_{\mathbf{d}}}{\sigma_{\mathbf{d}}}
\]

Where $\mu_{\mathbf{d}}$ and $\sigma_{\mathbf{d}}$ are the mean and standard deviation of the pairwise differences, respectively. The $\pmb{\hat{\Sigma}}\textsubscript{SPDC}$ is,

\begin{equation}
    \pmb{\hat{\Sigma}}\textsubscript{SPDC} = \frac{2}{n^2(n-1)}\mathbf{E}_{\text{S}}
\end{equation}

The objective of the local scaling approach is to evenly distribute the influence of different pairs by scaling the pairwise differences according to the local variance. This ensures that no single pairwise difference disproportionately affects the final covariance estimate. The regularization technique involves calculating local variances for each observation and scaling the pairwise differences by these local variances, defined as the variance of the pairwise differences for each observation relative to its neighbors:

\[
\text{scaled\_}\mathbf{D}_{ij} = \frac{\mathbf{d}_{ij}}{\sqrt{\text{var}(\mathbf{d}_{i,\cdot}) + \text{var}(\mathbf{d}_{j,\cdot})}}
\]

Where $\text{var}(\mathbf{d}_{i,\cdot})$ represents the variance of the pairwise differences involving observation $i$. The $\pmb{\hat{\Sigma}}\textsubscript{LSPDC}$ is,

    \begin{equation}
        \pmb{\hat{\Sigma}}\textsubscript{LSPDC} = \frac{2}{n^2(n-1)} \mathbf{E}_{\text{LS}}    
    \end{equation}

The MAXPDC approach aims to minimize the impact of outliers and extreme values on the covariance estimate. The regularization technique involves normalizing each pairwise difference such that its maximum absolute value is constrained:

\[
\text{scaled\_}\mathbf{D}_{ij} = \frac{\mathbf{d}_{ij}}{\max(|\mathbf{d}|)}
\]

where $\max(|\mathbf{d}|)$ represents the maximum absolute value of the pairwise differences. The $\pmb{\hat{\Sigma}}\textsubscript{MAXPDC}$ is,

    \begin{equation}
      \pmb{\hat{\Sigma}}\textsubscript{MAXPDC} = \frac{2}{n^2(n-1)}\mathbf{E}_{\text{MAX}}  
    \end{equation}

The RPDC approach aims to ensure a balanced representation across all pairwise differences by normalizing each difference by its range. The regularization technique involves scaling the pairwise differences by their range (the difference between the maximum and minimum values):

\[
\text{scaled\_}\mathbf{D}_{ij} = \frac{\mathbf{d}_{ij}}{\max(\mathbf{d}) - \min(\mathbf{d})}
\]

Where $\max(\mathbf{d})$ and $\min(\mathbf{d})$ represent the maximum and minimum values of the pairwise differences, respectively. The $\pmb{\hat{\Sigma}}\textsubscript{RPDC}$ is,

    \begin{equation}
        \pmb{\hat{\Sigma}}\textsubscript{RPDC} = \frac{2}{n^2(n-1)}\mathbf{E}_{\text{R}}
    \end{equation}

\section{Experiments on Synthetic Data}

As we discussed in the previous sections, the typical sample covariance estimation is ill-conditioned when \textit{n} $<$ \textit{p} high-dimensional data settings. It leads to misinterpreted and inaccurate PCA when \textit{n} $<$ \textit{p} high-dimensional data settings. Therefore, in this section, we explore the performance of our suggested PDC estimation and its regularized versions as a covariance estimation for PCA when \textit{n} $<$ \textit{p} high-dimensional data settings. We compare these suggested methods of covariance with traditional techniques such as MLE of covariance and Ledoit-Wolf covariance estimation under the context of PCA when \textit{n} $<$ \textit{p} high-dimensional data settings.

\subsection{Simulation Model}

To ensure reliable and robust simulation results, we assume \(\textbf{X}_i \sim \text{N}_p(\textbf{0}, \pmb{\Sigma})\). Here, \(\textbf{X}_i\) represents the \(i\)-th dimension, \(\textbf{0}\) means a \(p \times 1\) zero mean vector and \pmb{\(\Sigma\)} means a \(p \times p\) variance-covariance matrix. Here is an overview of the simulation settings, highlighting key aspects and their rationale.\\

\begin{enumerate}
    \item \textbf{Covariance Matrix (\pmb{$\Sigma$}):} This \(p \times p\) matrix contains random numbers drawn from a normal distribution with mean zero and variance one (i.e., (\(t = \text{randn}(p)\))). \pmb{\(\Sigma\)} = \(t \times t^T\), where \(t^T\) is the transpose of matrix \(t\). 

    \item \textbf{Mean Vector (\pmb{$\mu$}):} This \(p \times 1\) matrix/vector contains all zeros.

    \item \textbf{Number of Dimensions (\textit{p}):} \(p\) is set to 20 for ease of visualization.

    \item \textbf{Number of Observations (\textit{n}):} In this study, \(n\) is chosen such that \(n < p\). 

    \item \textbf{Number of Iterations (\textit{m}):} To ensure reliable simulation results, we conduct 500 iterations of each PCA with all the different \textit{n} $<$ \textit{p} high-dimensional data settings.
\end{enumerate}

\subsection{Performance Criterions}

To evaluate how well these different covariance estimation methods capture the underlying variance-covariance structure in terms of PCA, we used three different performance criterions as follows. 

\begin{enumerate}
    \item \textbf{Percentage of Variance Explained by Each Sample PC}: This criterion evaluates how much of the data's total variability is captured by each PC. The percentage of variance explained by each population PC is compared with that of the sample PCs to assess the accuracy of different PCA methods. Higher percentages indicate a better representation of data patterns by the PCs. By comparing these percentages, we can determine which PCA methods best capture the data's underlying patterns, aiding in selecting the most effective method for various analyses.\\
    
    \item \textbf{Cosine Similarity Error (CSE)}: This criterion measures the similarity between sample PCs and population PCs using the cosine of the angle between them. The formula for CSE is:
    
    \[
    CSE (PC_{i}) = 1 - \frac{{\hat{PC}_{i} \cdot PC_{i}}}{{\|\hat{PC}_{i}\| \times \|PC_{i}\|}}
    \]
    
    Here, \(\hat{PC_{i}}\) represents the \(i\)-th sample PC, and \(PC_{i}\) represents the \(i\)-th population PC. A smaller CSE indicates higher similarity, meaning the sample PCs are more accurately aligned with the population PCs. A higher CSE indicates lower similarity, meaning the sample PCs are less likely to align with the population PCs.\\

       \item \textbf{Minimizing the Objective Function}: As we discussed in earlier chapters, we know that the first \textit{n} -1 sample PCs may overestimate the population PCs when \textit{n} $<$ \textit{p} high-dimensional data settings. This is called the overdispersion of the sample PCA when \textit{n} $<$ \textit{p} high-dimensional data settings. In this research, we primarily focused on minimizing this issue of overdispersion of the sample PCA when \textit{n} $<$ \textit{p} high-dimensional data settings. Therefore, we used the following objective function to minimize the overdispersion of the sample PCA when \textit{n} $<$ \textit{p} high-dimensional data settings.  In mathematical terms, the objective function is to minimize:
    
    \[
    \text{Overdispersion} = \sum_{i=1}^p \left( \hat{\pi}_1 - \pi_1^{\text{pop}} \right)^2 \times \frac{p}{n-1}
    \]
    
    where \(\hat{\pi}_1\) is the proportion of variance explained by the first sample PC, and \(\pi_1^{\text{pop}}\) is the proportion of variance explained by the first population PC.

    For small \( n \), the sample eigenvalues \( \hat{\lambda}_i \) are less stable and exhibit more significant variability. However, the proportion of variance explained by the first sample PC significantly deviates from the proportion of variance explained by the first population PC.

Thus, the overdispersion:

\[
\text{Overdispersion} = \mathbb{E} \left[ \left( \hat{\pi}_1 - \pi_1^{\text{pop}} \right)^2 \right] \times \frac{p}{n-1} \text{ is large for small } n
\]

In other words, this function's value is low, indicating low overdispersion while ensuring the sample PCs closely align with the population PCs, and the function's value is high, indicating high overdispersion while providing the sample PCs not closely aligned with the population PCs.

\end{enumerate}

\subsection{Rationale and Relevance}

This study aims to identify reliable covariance estimation methods for PCA in \textit{n} $<$ \textit{p} high-dimensional data settings. For that purpose, in this study, we used datasets that follow a multivariate normal distribution with a random covariance matrix to ensure the results are generalizable and applicable to a wide range of real-world scenarios. Moreover, we use the first PC to compare the different PCA scenarios because the first PC plays a vital role in the PCA.

\subsection{PCA using PDC when \textit{n} \textless \textit{p}}

In previous section, we introduced the PDC estimation and four different regularized versions of PDC estimation for analyzing PCs when \textit{n} $<$ \textit{p} high-dimensional data settings. All of these methods aim to provide precise sample covariance estimation for \textit{n} $<$ \textit{p} high-dimensional data settings by taking all pairwise differences within the original dataset. The most highlighted thing about these estimators is that they enhance the sample size using the same number of observations we used in typical covariance estimations. However, the covariance matrices generated by the MLE and PDC methods exhibit distinct patterns in the data structure. For example, consider the case where \(p=20\) and \(n=5\) with data drawn from a multivariate normal distribution: \(\textbf{X}_i \sim \text{N}_p(\textbf{0}, \pmb{\Sigma})\). The closer inspection of Levene's test for variances results highlights that the diagonal and off-diagonal elements significantly contribute to the differences between the PDC and MLE covariance matrices. However, the off-diagonal elements show a higher Levene's test statistic and a lower p-value, indicating they contribute more significantly to the overall differences. This simple adjustment of the data matrix provides a slightly different scaled \pmb{S} (see Figure 1). However, the Frobenius norm values suggest that the MLE of the covariance matrix approximates true $\pmb{\Sigma}$ more closely than the PDC matrix. Specifically, the Frobenius norm for MLE (15.975) is lower than that for PDC (18.629), indicating that the MLE of the covariance matrix is closer to $\pmb{\Sigma}$.\\

\begin{table}[ht]
\centering
\begin{tabular}{@{}l@{}}
\toprule
\rowcolor[HTML]{C0C0C0} 
\textbf{Levene’s Test for Variances}  \\ \midrule
Levene’s test evaluates whether the variances of the variances and covariances           \\
between the two matrices are statistically significantly different.                      \\
{\ul \textbf{Diagonal Elements:}}     \\
Statistic: 22.986                     \\
p-value: 2.52e-05                     \\
Interpretation: The variances of the variances are significantly different (p < 0.05).   \\
{\ul \textbf{Off-Diagonal Elements:}} \\
Statistic: 249.528                    \\
p-value: 8.38e-49                     \\
Interpretation: The variances of the covariances are significantly different (p < 0.05). \\ \bottomrule
\end{tabular}
\end{table}

Although the MLE of the covariance provides an estimate closer to the true covariance matrix \(\pmb{\Sigma}\), the PDC method offers significant practical advantages. These benefits are particularly evident in PCA when dealing with high-dimensional data settings where \textit{n} $<$ \textit{p}.


\begin{figure}[!htbp] 
\centering
\includegraphics[width=1.0\textwidth,keepaspectratio]{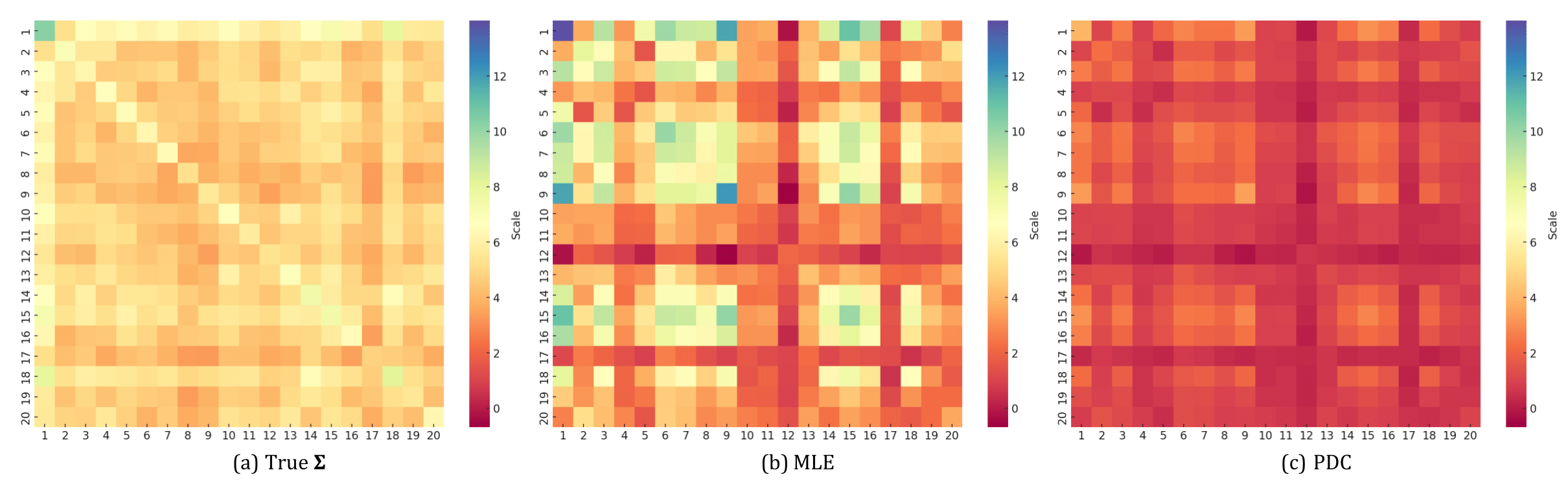}
\caption{Comparison of Sigma, MLE, and PDC}\label{fig4}
\end{figure}

Table 1 presents the average percentage of total explained variance for both the first \(n-1\) PCs and the last \(p-n+1\) PCs for a few selected high-dimensional data settings where \(n < p\). It demonstrates how MLE and PDC address overdispersion in the first \(n-1\) PCs and underdispersion in the last \(p-n+1\) PCs.\\

As shown in Table 1, PDC can reduce the overdispersion observed in the first \(n-1\) PCs marginally and to reduce the underdispersion observed in the last \(p-n+1\) PCs marginally. As per the Ledoit-Wolf method, PDC aims to achieve a more balanced variance distribution across all PCs by transforming significant variance from the first \(n-1\) PCs to the last \(p-n+1\) PCs. But it is not uniformly distributed. This ensures that PDC marginally (but significantly) improves the interpretability and reliability of PCA by minimizing the overdispersion and underdispersion of the sample PCA when \textit{n} $<$ \textit{p} high-dimensional data settings. However, the first PC is crucial in PCA as it captures the most variance in the data.

\pmb{Table 1:} Advantages of PDC Over MLE in high-dimensional Analysis
\begin{table}[ht]
\centering
\resizebox{\columnwidth}{!}{%
\begin{tabular}{@{}|ccllc|@{}}
\toprule
\rowcolor[HTML]{D9D9D9} 
\multicolumn{5}{|c|}{\cellcolor[HTML]{D9D9D9}\textbf{Average Percentage of Total Explained Variance of First (n-1) PCs}} \\ \midrule
\rowcolor[HTML]{D9D9D9} 
\multicolumn{1}{|c|}{\cellcolor[HTML]{D9D9D9}\textbf{p}} &
  \multicolumn{1}{c|}{\cellcolor[HTML]{D9D9D9}\textbf{n}} &
  \multicolumn{1}{c|}{\cellcolor[HTML]{D9D9D9}\textbf{MLE}} &
  \multicolumn{1}{c|}{\cellcolor[HTML]{D9D9D9}\textbf{PDC}} &
  \textbf{Improvement} \\ \midrule
\multicolumn{1}{|c|}{20} &
  \multicolumn{1}{c|}{7} &
  \multicolumn{1}{l|}{0.999999999999999793815041043000} &
  \multicolumn{1}{l|}{0.999999999999999520000000000000} &
  \cellcolor[HTML]{FFFFC7}\textbf{2.73815E-14 $\%$} \\ \midrule
\multicolumn{1}{|c|}{20} &
  \multicolumn{1}{c|}{10} &
  \multicolumn{1}{l|}{0.999999999999999842919079727000} &
  \multicolumn{1}{l|}{0.999999999999999768663111900000} &
  \cellcolor[HTML]{FFFFC7}\textbf{7.42559E-15 $\%$} \\ \midrule
\multicolumn{1}{|c|}{20} &
  \multicolumn{1}{c|}{12} &
  \multicolumn{1}{l|}{0.999999999999999998129935989200} &
  \multicolumn{1}{l|}{0.999999999999999837355171300000} &
  \cellcolor[HTML]{FFFFC7}\textbf{1.60775E-14 $\%$} \\ \midrule
\rowcolor[HTML]{D9D9D9} 
\multicolumn{5}{|c|}{\cellcolor[HTML]{D9D9D9}\textbf{Average Percentage of Total Explained Variance of Last (p-n+1) PCs}} \\ \midrule
\rowcolor[HTML]{D9D9D9} 
\multicolumn{1}{|c|}{\cellcolor[HTML]{D9D9D9}\textbf{p}} &
  \multicolumn{1}{c|}{\cellcolor[HTML]{D9D9D9}\textbf{n}} &
  \multicolumn{1}{c|}{\cellcolor[HTML]{D9D9D9}\textbf{MLE}} &
  \multicolumn{1}{c|}{\cellcolor[HTML]{D9D9D9}\textbf{PDC}} &
  \textbf{Improvement} \\ \midrule
\multicolumn{1}{|c|}{20} &
  \multicolumn{1}{c|}{7} &
  \multicolumn{1}{l|}{0.000000000000000206184958957000} &
  \multicolumn{1}{l|}{0.000000000000000480000000000000} &
  \cellcolor[HTML]{FFFFC7}\textbf{2.73815E-14 $\%$} \\ \midrule
\multicolumn{1}{|c|}{20} &
  \multicolumn{1}{c|}{10} &
  \multicolumn{1}{l|}{0.000000000000000157080920273000} &
  \multicolumn{1}{l|}{0.000000000000000231336888100000} &
  \cellcolor[HTML]{FFFFC7}\textbf{7.42559E-15 $\%$} \\ \midrule
\multicolumn{1}{|c|}{20} &
  \multicolumn{1}{c|}{12} &
  \multicolumn{1}{l|}{0.000000000000000001870064010800} &
  \multicolumn{1}{l|}{0.000000000000000162644828700000} &
  \cellcolor[HTML]{FFFFC7}\textbf{1.60775E-14 $\%$} \\ \bottomrule
\end{tabular}%
}
\end{table}

 Therefore, in Table 2, we specifically examine cases where \(n < p\) to explore the impact of PDC on the first PC. This analysis helps us understand how PDC enhances the finite sample PC estimates in high-dimensional settings when \textit{n} $<$ \textit{p}. By doing so, we can see how PDC improves the accuracy and utility of PCA in these contexts.\\

\pmb{Table 2:} Average PDC Insights for the first PC
\begin{table}[ht]
\centering
\resizebox{\columnwidth}{!}{%
\begin{tabular}{@{}|l|l|cr
>{\columncolor[HTML]{FFFFC7}}r |@{}}
\toprule
\multicolumn{1}{|c|}{\cellcolor[HTML]{D9D9D9}} &
  \multicolumn{1}{c|}{\cellcolor[HTML]{D9D9D9}} &
  \multicolumn{3}{c|}{\cellcolor[HTML]{D9D9D9}\textbf{Average Overdispersion of first PC}} \\ \cmidrule(l){3-5} 
\multicolumn{1}{|c|}{\multirow{-2}{*}{\cellcolor[HTML]{D9D9D9}\textbf{p}}} &
  \multicolumn{1}{c|}{\multirow{-2}{*}{\cellcolor[HTML]{D9D9D9}\textbf{n}}} &
  \multicolumn{1}{c|}{\cellcolor[HTML]{D9D9D9}\textbf{Pop}} &
  \multicolumn{1}{c|}{\cellcolor[HTML]{D9D9D9}\textbf{MLE}} &
  \multicolumn{1}{c|}{\cellcolor[HTML]{D9D9D9}\textbf{PDC}} \\ \midrule
20 &
  7 &
  \multicolumn{1}{c|}{0} &
  \multicolumn{1}{r|}{0.004946670881132785892175807} &
  0.0049466708811327\textbf{286463011} \\ \midrule
20 &
  10 &
  \multicolumn{1}{c|}{0} &
  \multicolumn{1}{r|}{0.001897650500155690376016815} &
  0.0018976505001556\textbf{47007929915} \\ \midrule
20 &
  12 &
  \multicolumn{1}{c|}{0} &
  \multicolumn{1}{r|}{0.0005406391043933448650390949} &
  0.0005406391043933\textbf{37926145191} \\ \midrule
20 &
  16 &
  \multicolumn{1}{c|}{0} &
  \multicolumn{1}{r|}{0.0005308689153643464820470088} &
  0.0005308689153643\textbf{345558231115} \\ \midrule
20 &
  17 &
  \multicolumn{1}{c|}{0} &
  \multicolumn{1}{r|}{0.0002563863721771374648214503} &
  0.00025638637217713\textbf{35074835207} \\ \midrule
\multicolumn{1}{|c|}{\cellcolor[HTML]{D9D9D9}} &
  \multicolumn{1}{c|}{\cellcolor[HTML]{D9D9D9}} &
  \multicolumn{3}{c|}{\cellcolor[HTML]{D9D9D9}\textbf{Average Percentage of Explained Variance of first PC}} \\ \cmidrule(l){3-5} 
\multicolumn{1}{|c|}{\multirow{-2}{*}{\cellcolor[HTML]{D9D9D9}\textbf{p}}} &
  \multicolumn{1}{c|}{\multirow{-2}{*}{\cellcolor[HTML]{D9D9D9}\textbf{n}}} &
  \multicolumn{1}{c|}{\cellcolor[HTML]{D9D9D9}\textbf{Pop}} &
  \multicolumn{1}{c|}{\cellcolor[HTML]{D9D9D9}\textbf{MLE}} &
  \multicolumn{1}{c|}{\cellcolor[HTML]{D9D9D9}\textbf{PDC}} \\ \midrule
20 &
  7 &
  \multicolumn{1}{r|}{76.50551356458754526101984} &
  \multicolumn{1}{r|}{80.3577872624701825543525} &
  80.357787262470\textbf{15413264307} \\ \midrule
20 &
  10 &
  \multicolumn{1}{r|}{76.50551356458754526101984} &
  \multicolumn{1}{r|}{79.42774340679719102809031} &
  79.4277434067971\textbf{6260638088} \\ \midrule
20 &
  12 &
  \multicolumn{1}{r|}{76.50551356458754526101984} &
  \multicolumn{1}{r|}{78.22990188343089812406106} &
  78.2299018834308\textbf{8391320634} \\ \midrule
20 &
  16 &
  \multicolumn{1}{r|}{76.50551356458754526101984} &
  \multicolumn{1}{r|}{78.50088743061681384460826} &
  78.500887430616\textbf{79963375354} \\ \midrule
20 &
  17 &
  \multicolumn{1}{r|}{76.50551356458754526101984} &
  \multicolumn{1}{r|}{77.93767660608006053735153} &
  77.9376766060800\textbf{4632649681} \\ \midrule
\multicolumn{1}{|c|}{\cellcolor[HTML]{D9D9D9}} &
  \multicolumn{1}{c|}{\cellcolor[HTML]{D9D9D9}} &
  \multicolumn{3}{c|}{\cellcolor[HTML]{D9D9D9}\textbf{Average CSE of first PC}} \\ \cmidrule(l){3-5} 
\multicolumn{1}{|c|}{\multirow{-2}{*}{\cellcolor[HTML]{D9D9D9}\textbf{p}}} &
  \multicolumn{1}{c|}{\multirow{-2}{*}{\cellcolor[HTML]{D9D9D9}\textbf{n}}} &
  \multicolumn{1}{c|}{\cellcolor[HTML]{D9D9D9}\textbf{Pop}} &
  \multicolumn{1}{c|}{\cellcolor[HTML]{D9D9D9}\textbf{MLE}} &
  \multicolumn{1}{c|}{\cellcolor[HTML]{D9D9D9}\textbf{PDC}} \\ \midrule
20 &
  7 &
  \multicolumn{1}{c|}{0} &
  \multicolumn{1}{r|}{0.04886216868271503477316742} &
  0.0488621686827150\textbf{070175918} \\ \midrule
20 &
  10 &
  \multicolumn{1}{c|}{0} &
  \multicolumn{1}{r|}{0.02187722517434496916077791} &
  0.02187722517434496\textbf{569133096} \\ \midrule
20 &
  12 &
  \multicolumn{1}{c|}{0} &
  \multicolumn{1}{r|}{0.01885453113147821796458103} &
  0.01885453113147821\textbf{102568713} \\ \midrule
20 &
  16 &
  \multicolumn{1}{c|}{0} &
  \multicolumn{1}{r|}{0.01222599678332552362858276} &
  0.01222599678332551\textbf{668968886} \\ \midrule
20 &
  17 &
  \multicolumn{1}{c|}{0} &
  \multicolumn{1}{r|}{0.01138314510337739768330856} &
  \multicolumn{1}{r|}{0.01138314510337739768330856} \\ \bottomrule
\end{tabular}%
}
\end{table}

As we highlighted previously, the first sample PC shows overdispersed estimates when \textit{n} $<$ \textit{p} high-dimensional data settings. That means the percentage of variance explained by the estimated first PC is artificially high, with some noise other than the actual variance. However, according to Table 2, PDC consistently reduces the overdispersion of the first PC marginally and is closer to the actual population PC than MLE. These improvements, though marginally low (on the order of \(10^{-15}\)), provide a more balanced variance distribution and enhance PCA performance. The enhancements are too minor but persistent compared to traditional methods. This tendency can be explained by the principles of perturbation theory and sensitivity analysis outlined previously. According to that, the minor changes in input parameters (such as \textit{n}) lead to predictable and marginally different outcomes in covariance estimation.\\

Moreover, Table 2 shows the distribution of the average CSE of the first PC across different high-dimensional data settings where \textit{n} $<$ \textit{p}. According to that, as we have experienced before, PDC gives the lowest CSE than MLE while ensuring that PCs are more closely aligned with the population of PCs than MLE.\\

The above results ensure that, in the context of PCA for high-dimensional data settings where \textit{n} $<$ \textit{p}, even a marginal improvement in covariance estimation can significantly influence the final output. Numerically, it is too small. However, marginal improvements can yield significant performance in fields like genomics, finance, health science, etc.

\section{PCA using Regularized PDCs}

When $n < p$, we can’t obtain more than $n$ meaningful PCs. The first few PCs in such cases are often overdispersed. To compute $p$ meaningful PCs without increasing the sample size, which might be expensive or impractical, we can force the distribution of eigenvalues of the sample covariance matrix, as demonstrated by Ledoit and Wolf. According to the state of the art, while there are many alternatives for this approach (though not specifically based on PCA), Ledoit-Wolf covariance estimation is the most renowned and effective method for high-dimensional analysis when $n < p$. However, in the context of PCA for high-dimensional settings with $n < p$, Ledoit-Wolf does not significantly outperform the traditional MLE. Its uniform shrinkage of the last $p-n+1$ eigenvalues leads to an unnecessary overestimation of these PCs and causes substantial underdispersion of the first PC. This ensures that the last $p-n+1$ eigenvalues do not approach zero or nearly zero values, drastically reducing the first PC. Even if the first PC is crucial in PCA, according to Ledoit Wolf covariance method does not capture the actual amount of variance in the data due to Ledoit-Wolf shrinking the eigenvalues excessively and causing under dispersion of first \textit{n} -1 PCs and loss of valuable information. The unnecessary uniform distribution of last \textit{p}-\textit{n}+1 PCs may affect this. However, despite its drawbacks, MLE is preferred to preserve the importance of the first PC. That’s why we still use MLE for the covariance estimation \textit{n} $<$ \textit{p} high-dimensional data settings as well. Considering all these issues, our proposed covariance estimation method PDC balances eigenvalues effectively without increasing the sample size, showing slight improvements in PCA estimates and addressing the limitations of traditional methods when \textit{n} $<$ \textit{p} high-dimensional data settings. Given the limited improvement in PDC estimation, in section 04, we proposed four different regularized PDCs to enhance their performance. In the following sections, we will evaluate the performance of these regularizations to determine their effectiveness.\\

\pmb{Table 3:} Average Overdispersion of the first PC
\begin{table}[ht]
\centering
\resizebox{\columnwidth}{!}{%
\begin{tabular}{@{}|c|c|cccc
>{\columncolor[HTML]{CBCEFB}}c cc
>{\columncolor[HTML]{9AFF99}}c |@{}}
\toprule
\cellcolor[HTML]{D9D9D9} &
  \cellcolor[HTML]{D9D9D9} &
  \multicolumn{8}{c|}{\cellcolor[HTML]{D9D9D9}\textbf{Average Overdispersion of first PC}} \\ \cmidrule(l){3-10} 
\multirow{-2}{*}{\cellcolor[HTML]{D9D9D9}\textbf{p}} &
  \multirow{-2}{*}{\cellcolor[HTML]{D9D9D9}\textbf{n}} &
  \multicolumn{1}{c|}{\cellcolor[HTML]{D9D9D9}\textbf{POP}} &
  \multicolumn{1}{c|}{\cellcolor[HTML]{D9D9D9}\textbf{MLE}} &
  \multicolumn{1}{c|}{\cellcolor[HTML]{D9D9D9}\textbf{LW}} &
  \multicolumn{1}{c|}{\cellcolor[HTML]{D9D9D9}\textbf{PDC}} &
  \multicolumn{1}{c|}{\cellcolor[HTML]{D9D9D9}\textbf{SPDC}} &
  \multicolumn{1}{c|}{\cellcolor[HTML]{D9D9D9}\textbf{LSPDC}} &
  \multicolumn{1}{c|}{\cellcolor[HTML]{D9D9D9}\textbf{MAXPDC}} &
  \cellcolor[HTML]{D9D9D9}\textbf{RPDC} \\ \midrule
20 &
  3 &
  \multicolumn{1}{c|}{0.000000} &
  \multicolumn{1}{c|}{0.149947} &
  \multicolumn{1}{c|}{0.037591} &
  \multicolumn{1}{c|}{0.149947} &
  \multicolumn{1}{c|}{\cellcolor[HTML]{9AFF99}0.005540} &
  \multicolumn{1}{c|}{\cellcolor[HTML]{FFFC9E}0.003689} &
  \multicolumn{1}{c|}{0.066781} &
  \cellcolor[HTML]{CBCEFB}0.005602 \\ \midrule
20 &
  4 &
  \multicolumn{1}{c|}{0.000000} &
  \multicolumn{1}{c|}{0.040505} &
  \multicolumn{1}{c|}{0.140737} &
  \multicolumn{1}{c|}{0.040505} &
  \multicolumn{1}{c|}{\cellcolor[HTML]{9AFF99}0.001819} &
  \multicolumn{1}{c|}{\cellcolor[HTML]{CBCEFB}0.001936} &
  \multicolumn{1}{c|}{0.008969} &
  \cellcolor[HTML]{FFFC9E}0.001664 \\ \midrule
20 &
  5 &
  \multicolumn{1}{c|}{0.000000} &
  \multicolumn{1}{c|}{0.017546} &
  \multicolumn{1}{c|}{0.126138} &
  \multicolumn{1}{c|}{0.017546} &
  \multicolumn{1}{c|}{\cellcolor[HTML]{9AFF99}0.001741} &
  \multicolumn{1}{c|}{\cellcolor[HTML]{FFFC9E}0.001553} &
  \multicolumn{1}{c|}{0.002676} &
  \cellcolor[HTML]{CBCEFB}0.001779 \\ \midrule
20 &
  6 &
  \multicolumn{1}{c|}{0.000000} &
  \multicolumn{1}{c|}{0.008383} &
  \multicolumn{1}{c|}{0.096425} &
  \multicolumn{1}{c|}{0.008383} &
  \multicolumn{1}{c|}{\cellcolor[HTML]{CBCEFB}0.001780} &
  \multicolumn{1}{c|}{0.002146} &
  \multicolumn{1}{c|}{\cellcolor[HTML]{FFFC9E}0.000693} &
  0.001496 \\ \midrule
20 &
  7 &
  \multicolumn{1}{c|}{0.000000} &
  \multicolumn{1}{c|}{0.004947} &
  \multicolumn{1}{c|}{0.076989} &
  \multicolumn{1}{c|}{0.004947} &
  \multicolumn{1}{c|}{\cellcolor[HTML]{CBCEFB}0.001133} &
  \multicolumn{1}{c|}{0.001174} &
  \multicolumn{1}{c|}{\cellcolor[HTML]{FFFC9E}0.000430} &
  0.000996 \\ \midrule
20 &
  8 &
  \multicolumn{1}{c|}{0.000000} &
  \multicolumn{1}{c|}{0.004130} &
  \multicolumn{1}{c|}{0.053602} &
  \multicolumn{1}{c|}{0.004130} &
  \multicolumn{1}{c|}{\cellcolor[HTML]{9AFF99}0.000349} &
  \multicolumn{1}{c|}{\cellcolor[HTML]{CBCEFB}0.000380} &
  \multicolumn{1}{c|}{0.000406} &
  \cellcolor[HTML]{FFFC9E}0.000270 \\ \midrule
20 &
  9 &
  \multicolumn{1}{c|}{0.000000} &
  \multicolumn{1}{c|}{0.002951} &
  \multicolumn{1}{c|}{0.041354} &
  \multicolumn{1}{c|}{0.002951} &
  \multicolumn{1}{c|}{\cellcolor[HTML]{9AFF99}0.000232} &
  \multicolumn{1}{c|}{\cellcolor[HTML]{CBCEFB}0.000266} &
  \multicolumn{1}{c|}{0.000357} &
  \cellcolor[HTML]{FFFC9E}0.000195 \\ \midrule
20 &
  10 &
  \multicolumn{1}{c|}{0.000000} &
  \multicolumn{1}{c|}{0.001898} &
  \multicolumn{1}{c|}{0.036120} &
  \multicolumn{1}{c|}{0.001898} &
  \multicolumn{1}{c|}{\cellcolor[HTML]{CBCEFB}0.000217} &
  \multicolumn{1}{c|}{0.000231} &
  \multicolumn{1}{c|}{\cellcolor[HTML]{FFFC9E}0.000118} &
  0.000179 \\ \midrule
20 &
  11 &
  \multicolumn{1}{c|}{0.000000} &
  \multicolumn{1}{c|}{0.001543} &
  \multicolumn{1}{c|}{0.027566} &
  \multicolumn{1}{c|}{0.001543} &
  \multicolumn{1}{c|}{\cellcolor[HTML]{CBCEFB}0.000261} &
  \multicolumn{1}{c|}{0.000273} &
  \multicolumn{1}{c|}{\cellcolor[HTML]{FFFC9E}0.000093} &
  0.000225 \\ \midrule
20 &
  12 &
  \multicolumn{1}{c|}{0.000000} &
  \multicolumn{1}{c|}{0.000541} &
  \multicolumn{1}{c|}{0.026169} &
  \multicolumn{1}{c|}{0.000541} &
  \multicolumn{1}{c|}{\cellcolor[HTML]{CBCEFB}0.000696} &
  \multicolumn{1}{c|}{0.000718} &
  \multicolumn{1}{c|}{\cellcolor[HTML]{FFFC9E}0.000018} &
  0.000658 \\ \midrule
20 &
  13 &
  \multicolumn{1}{c|}{0.000000} &
  \multicolumn{1}{c|}{0.000900} &
  \multicolumn{1}{c|}{0.018428} &
  \multicolumn{1}{c|}{0.000900} &
  \multicolumn{1}{c|}{\cellcolor[HTML]{9AFF99}0.000084} &
  \multicolumn{1}{c|}{\cellcolor[HTML]{CBCEFB}0.000087} &
  \multicolumn{1}{c|}{0.000097} &
  \cellcolor[HTML]{FFFC9E}0.000061 \\ \midrule
20 &
  14 &
  \multicolumn{1}{c|}{0.000000} &
  \multicolumn{1}{c|}{0.000366} &
  \multicolumn{1}{c|}{0.016601} &
  \multicolumn{1}{c|}{0.000366} &
  \multicolumn{1}{c|}{\cellcolor[HTML]{CBCEFB}0.000269} &
  \multicolumn{1}{c|}{0.000292} &
  \multicolumn{1}{c|}{\cellcolor[HTML]{FFFC9E}0.000008} &
  0.000211 \\ \midrule
20 &
  15 &
  \multicolumn{1}{c|}{0.000000} &
  \multicolumn{1}{c|}{0.000410} &
  \multicolumn{1}{c|}{0.013254} &
  \multicolumn{1}{c|}{0.000410} &
  \multicolumn{1}{c|}{\cellcolor[HTML]{CBCEFB}0.000152} &
  \multicolumn{1}{c|}{0.000163} &
  \multicolumn{1}{c|}{\cellcolor[HTML]{FFFC9E}0.000014} &
  0.000137 \\ \midrule
20 &
  16 &
  \multicolumn{1}{c|}{0.000000} &
  \multicolumn{1}{c|}{0.000531} &
  \multicolumn{1}{c|}{0.010264} &
  \multicolumn{1}{c|}{0.000531} &
  \multicolumn{1}{c|}{\cellcolor[HTML]{CBCEFB}0.000047} &
  \multicolumn{1}{c|}{0.000057} &
  \multicolumn{1}{c|}{\cellcolor[HTML]{9AFF99}0.000032} &
  \cellcolor[HTML]{FFFC9E}0.000031 \\ \midrule
20 &
  17 &
  \multicolumn{1}{c|}{0.000000} &
  \multicolumn{1}{c|}{0.000256} &
  \multicolumn{1}{c|}{0.010572} &
  \multicolumn{1}{c|}{0.000256} &
  \multicolumn{1}{c|}{\cellcolor[HTML]{CBCEFB}0.000107} &
  \multicolumn{1}{c|}{0.000107} &
  \multicolumn{1}{c|}{\cellcolor[HTML]{FFFC9E}0.000005} &
  0.000093 \\ \midrule
20 &
  18 &
  \multicolumn{1}{c|}{0.000000} &
  \multicolumn{1}{c|}{0.000262} &
  \multicolumn{1}{c|}{0.007587} &
  \multicolumn{1}{c|}{0.000262} &
  \multicolumn{1}{c|}{\cellcolor[HTML]{CBCEFB}0.000051} &
  \multicolumn{1}{c|}{0.000065} &
  \multicolumn{1}{c|}{\cellcolor[HTML]{FFFC9E}0.000018} &
  0.000030 \\ \midrule
20 &
  19 &
  \multicolumn{1}{c|}{0.000000} &
  \multicolumn{1}{c|}{0.000243} &
  \multicolumn{1}{c|}{0.006996} &
  \multicolumn{1}{c|}{0.000243} &
  \multicolumn{1}{c|}{\cellcolor[HTML]{CBCEFB}0.000043} &
  \multicolumn{1}{c|}{0.000047} &
  \multicolumn{1}{c|}{\cellcolor[HTML]{FFFC9E}0.000018} &
  0.000034 \\ \midrule
20 &
  20 &
  \multicolumn{1}{c|}{0.000000} &
  \multicolumn{1}{c|}{0.000260} &
  \multicolumn{1}{c|}{0.005862} &
  \multicolumn{1}{c|}{0.000260} &
  \multicolumn{1}{c|}{\cellcolor[HTML]{CBCEFB}0.000038} &
  \multicolumn{1}{c|}{0.000041} &
  \multicolumn{1}{c|}{\cellcolor[HTML]{FFFC9E}0.000016} &
  0.000030 \\ \bottomrule
\end{tabular}%
}

\begin{tabular}{l @{\hspace{15pt}} c @{\hspace{15pt}} l @{\hspace{30pt}} c @{\hspace{15pt}} l @{\hspace{30pt}} c @{\hspace{15pt}} l}
\textbf{Note:} &
\cellcolor{yellow}{\hspace{0.01cm}\vspace{0.01cm}} & Best &
\cellcolor{green}{\hspace{0.01cm}\vspace{0.01cm}} & Second Best &
\cellcolor[HTML]{CBCEFB}{\hspace{0.01cm}\vspace{0.01cm}} & Third Best \\ \hline
\end{tabular}
\end{table}

Table 3 shows that the lowest average overdispersion of the first PC is achieved by MAXPDC, followed by RPDC and SPDC, respectively. Compared to the standard MLE method, all these regularized PDC methods significantly reduce the overdispersion of the first PC. This reduction in overdispersion becomes even more pronounced as the number of observations \(n\) decreases relative to the number of dimensions \(p\). The findings also indicate that Ledoit-Wolf covariance estimation is ineffective for PCA in high-dimensional settings where \(n < p\). This supports our previous assertion that we continue to use MLE despite the existence of alternative covariance estimation methods. These existing alternatives, while innovative, offer a different level of reliability in preserving the importance of the first PC in such high-dimensional scenarios.

\newpage

Table 4 reveals a similar pattern of results. Specifically, to those observed in Table 3, Table 4 shows the average percentage of explained variance of the first PC. As previously mentioned, the first PC is crucial in PCA. If the first PC is excessively overdispersed or underdispersed, it negatively impacts the PCA results.\\

\pmb{Table 4:} Average Percentage of Explained Variance of the first PC
\begin{table}[ht]
\centering
\resizebox{\columnwidth}{!}{%
\begin{tabular}{@{}|c|c|cccc
>{\columncolor[HTML]{CBCEFB}}c cc
>{\columncolor[HTML]{9AFF99}}c |@{}}
\toprule
\cellcolor[HTML]{D9D9D9} &
  \cellcolor[HTML]{D9D9D9} &
  \multicolumn{8}{c|}{\cellcolor[HTML]{D9D9D9}\textbf{Average Percentage of Explained Variance of first PC}} \\ \cmidrule(l){3-10} 
\multirow{-2}{*}{\cellcolor[HTML]{D9D9D9}\textbf{p}} &
  \multirow{-2}{*}{\cellcolor[HTML]{D9D9D9}\textbf{n}} &
  \multicolumn{1}{c|}{\cellcolor[HTML]{D9D9D9}\textbf{POP}} &
  \multicolumn{1}{c|}{\cellcolor[HTML]{D9D9D9}\textbf{MLE}} &
  \multicolumn{1}{c|}{\cellcolor[HTML]{D9D9D9}\textbf{LW}} &
  \multicolumn{1}{c|}{\cellcolor[HTML]{D9D9D9}\textbf{PDC}} &
  \multicolumn{1}{c|}{\cellcolor[HTML]{D9D9D9}\textbf{SPDC}} &
  \multicolumn{1}{c|}{\cellcolor[HTML]{D9D9D9}\textbf{LSPDC}} &
  \multicolumn{1}{c|}{\cellcolor[HTML]{D9D9D9}\textbf{MAXPDC}} &
  \cellcolor[HTML]{D9D9D9}\textbf{RPDC} \\ \midrule
20 &
  3 &
  \multicolumn{1}{c|}{76.50551} &
  \multicolumn{1}{c|}{88.75079} &
  \multicolumn{1}{c|}{70.37435} &
  \multicolumn{1}{c|}{88.75079} &
  \multicolumn{1}{c|}{\cellcolor[HTML]{9AFF99}78.85917} &
  \multicolumn{1}{c|}{\cellcolor[HTML]{FFFC9E}78.42632} &
  \multicolumn{1}{c|}{84.67747} &
  \cellcolor[HTML]{CBCEFB}78.87246 \\ \midrule
20 &
  4 &
  \multicolumn{1}{c|}{76.50551} &
  \multicolumn{1}{c|}{84.30026} &
  \multicolumn{1}{c|}{61.97607} &
  \multicolumn{1}{c|}{84.30026} &
  \multicolumn{1}{c|}{\cellcolor[HTML]{9AFF99}74.85354} &
  \multicolumn{1}{c|}{\cellcolor[HTML]{CBCEFB}74.80138} &
  \multicolumn{1}{c|}{80.17347} &
  \cellcolor[HTML]{FFFC9E}74.92578 \\ \midrule
20 &
  5 &
  \multicolumn{1}{c|}{76.50551} &
  \multicolumn{1}{c|}{82.42940} &
  \multicolumn{1}{c|}{60.62233} &
  \multicolumn{1}{c|}{82.42940} &
  \multicolumn{1}{c|}{\cellcolor[HTML]{9AFF99}74.63940} &
  \multicolumn{1}{c|}{\cellcolor[HTML]{FFFC9E}74.74337} &
  \multicolumn{1}{c|}{78.81875} &
  \cellcolor[HTML]{CBCEFB}74.61925 \\ \midrule
20 &
  6 &
  \multicolumn{1}{c|}{76.50551} &
  \multicolumn{1}{c|}{81.08342} &
  \multicolumn{1}{c|}{60.97936} &
  \multicolumn{1}{c|}{81.08342} &
  \multicolumn{1}{c|}{\cellcolor[HTML]{CBCEFB}74.39629} &
  \multicolumn{1}{c|}{74.18939} &
  \multicolumn{1}{c|}{\cellcolor[HTML]{FFFC9E}77.82131} &
  74.57150 \\ \midrule
20 &
  7 &
  \multicolumn{1}{c|}{76.50551} &
  \multicolumn{1}{c|}{80.35779} &
  \multicolumn{1}{c|}{61.30794} &
  \multicolumn{1}{c|}{80.35779} &
  \multicolumn{1}{c|}{\cellcolor[HTML]{CBCEFB}74.66194} &
  \multicolumn{1}{c|}{74.62862} &
  \multicolumn{1}{c|}{\cellcolor[HTML]{FFFC9E}77.64125} &
  74.77719 \\ \midrule
20 &
  8 &
  \multicolumn{1}{c|}{76.50551} &
  \multicolumn{1}{c|}{80.30731} &
  \multicolumn{1}{c|}{62.80860} &
  \multicolumn{1}{c|}{80.30731} &
  \multicolumn{1}{c|}{\cellcolor[HTML]{9AFF99}75.40000} &
  \multicolumn{1}{c|}{\cellcolor[HTML]{CBCEFB}75.35222} &
  \multicolumn{1}{c|}{77.69829} &
  \cellcolor[HTML]{FFFC9E}75.53381 \\ \midrule
20 &
  9 &
  \multicolumn{1}{c|}{76.50551} &
  \multicolumn{1}{c|}{79.94109} &
  \multicolumn{1}{c|}{63.64406} &
  \multicolumn{1}{c|}{79.94109} &
  \multicolumn{1}{c|}{\cellcolor[HTML]{9AFF99}75.54245} &
  \multicolumn{1}{c|}{\cellcolor[HTML]{CBCEFB}75.47423} &
  \multicolumn{1}{c|}{77.70081} &
  \cellcolor[HTML]{FFFC9E}75.62140 \\ \midrule
20 &
  10 &
  \multicolumn{1}{c|}{76.50551} &
  \multicolumn{1}{c|}{79.42774} &
  \multicolumn{1}{c|}{63.75641} &
  \multicolumn{1}{c|}{79.42774} &
  \multicolumn{1}{c|}{\cellcolor[HTML]{CBCEFB}75.51785} &
  \multicolumn{1}{c|}{75.48604} &
  \multicolumn{1}{c|}{\cellcolor[HTML]{FFFC9E}77.23406} &
  75.60792 \\ \midrule
20 &
  11 &
  \multicolumn{1}{c|}{76.50551} &
  \multicolumn{1}{c|}{79.28301} &
  \multicolumn{1}{c|}{64.76541} &
  \multicolumn{1}{c|}{79.28301} &
  \multicolumn{1}{c|}{\cellcolor[HTML]{CBCEFB}75.36271} &
  \multicolumn{1}{c|}{75.33792} &
  \multicolumn{1}{c|}{\cellcolor[HTML]{FFFC9E}77.18822} &
  75.44423 \\ \midrule
20 &
  12 &
  \multicolumn{1}{c|}{76.50551} &
  \multicolumn{1}{c|}{78.22990} &
  \multicolumn{1}{c|}{64.50851} &
  \multicolumn{1}{c|}{78.22990} &
  \multicolumn{1}{c|}{\cellcolor[HTML]{CBCEFB}74.54889} &
  \multicolumn{1}{c|}{74.51772} &
  \multicolumn{1}{c|}{\cellcolor[HTML]{FFFC9E}76.18879} &
  74.60364 \\ \midrule
20 &
  13 &
  \multicolumn{1}{c|}{76.50551} &
  \multicolumn{1}{c|}{78.82873} &
  \multicolumn{1}{c|}{65.99027} &
  \multicolumn{1}{c|}{78.82873} &
  \multicolumn{1}{c|}{\cellcolor[HTML]{9AFF99}75.79672} &
  \multicolumn{1}{c|}{\cellcolor[HTML]{CBCEFB}75.78465} &
  \multicolumn{1}{c|}{77.26744} &
  \cellcolor[HTML]{FFFC9E}75.90099 \\ \midrule
20 &
  14 &
  \multicolumn{1}{c|}{76.50551} &
  \multicolumn{1}{c|}{78.04745} &
  \multicolumn{1}{c|}{66.11768} &
  \multicolumn{1}{c|}{78.04745} &
  \multicolumn{1}{c|}{\cellcolor[HTML]{CBCEFB}75.18417} &
  \multicolumn{1}{c|}{75.12766} &
  \multicolumn{1}{c|}{\cellcolor[HTML]{FFFC9E}76.72951} &
  75.33387 \\ \midrule
20 &
  15 &
  \multicolumn{1}{c|}{76.50551} &
  \multicolumn{1}{c|}{78.19866} &
  \multicolumn{1}{c|}{66.87325} &
  \multicolumn{1}{c|}{78.19866} &
  \multicolumn{1}{c|}{\cellcolor[HTML]{CBCEFB}75.47357} &
  \multicolumn{1}{c|}{75.43793} &
  \multicolumn{1}{c|}{\cellcolor[HTML]{FFFC9E}76.81705} &
  75.52452 \\ \midrule
20 &
  16 &
  \multicolumn{1}{c|}{76.50551} &
  \multicolumn{1}{c|}{78.50089} &
  \multicolumn{1}{c|}{67.73184} &
  \multicolumn{1}{c|}{78.50089} &
  \multicolumn{1}{c|}{\cellcolor[HTML]{CBCEFB}75.91402} &
  \multicolumn{1}{c|}{75.85280} &
  \multicolumn{1}{c|}{\cellcolor[HTML]{9AFF99}76.99270} &
  \cellcolor[HTML]{FFFC9E}76.01952 \\ \midrule
20 &
  17 &
  \multicolumn{1}{c|}{76.50551} &
  \multicolumn{1}{c|}{77.93768} &
  \multicolumn{1}{c|}{67.30892} &
  \multicolumn{1}{c|}{77.93768} &
  \multicolumn{1}{c|}{\cellcolor[HTML]{CBCEFB}75.57830} &
  \multicolumn{1}{c|}{75.57884} &
  \multicolumn{1}{c|}{\cellcolor[HTML]{FFFC9E}76.71394} &
  75.64455 \\ \midrule
20 &
  18 &
  \multicolumn{1}{c|}{76.50551} &
  \multicolumn{1}{c|}{77.99916} &
  \multicolumn{1}{c|}{68.47479} &
  \multicolumn{1}{c|}{77.99916} &
  \multicolumn{1}{c|}{\cellcolor[HTML]{CBCEFB}75.84484} &
  \multicolumn{1}{c|}{75.76460} &
  \multicolumn{1}{c|}{\cellcolor[HTML]{FFFC9E}76.89616} &
  75.99961 \\ \midrule
20 &
  19 &
  \multicolumn{1}{c|}{76.50551} &
  \multicolumn{1}{c|}{77.98393} &
  \multicolumn{1}{c|}{68.57079} &
  \multicolumn{1}{c|}{77.98393} &
  \multicolumn{1}{c|}{\cellcolor[HTML]{CBCEFB}75.88555} &
  \multicolumn{1}{c|}{75.85635} &
  \multicolumn{1}{c|}{\cellcolor[HTML]{FFFC9E}76.90846} &
  75.94862 \\ \midrule
20 &
  20 &
  \multicolumn{1}{c|}{76.50551} &
  \multicolumn{1}{c|}{78.07642} &
  \multicolumn{1}{c|}{69.04307} &
  \multicolumn{1}{c|}{78.07642} &
  \multicolumn{1}{c|}{\cellcolor[HTML]{CBCEFB}75.90318} &
  \multicolumn{1}{c|}{75.88215} &
  \multicolumn{1}{c|}{\cellcolor[HTML]{FFFC9E}76.89435} &
  75.96801 \\ \bottomrule
\end{tabular}%
}

\begin{tabular}{l @{\hspace{15pt}} c @{\hspace{15pt}} l @{\hspace{30pt}} c @{\hspace{15pt}} l @{\hspace{30pt}} c @{\hspace{15pt}} l}
\textbf{Note:} &
\cellcolor{yellow}{\hspace{0.01cm}\vspace{0.01cm}} & Best &
\cellcolor{green}{\hspace{0.01cm}\vspace{0.01cm}} & Second Best &
\cellcolor[HTML]{CBCEFB}{\hspace{0.01cm}\vspace{0.01cm}} & Third Best \\ \hline
\end{tabular}
\end{table}

When examining the average percentage of explained variance of the first PC across different techniques, it is evident that the regularized versions of PDC provide much closer finite sample estimates for the first PC than the MLE method. Among these, MAXPDC offers the closest estimates for the first PC, followed by RPDC and SPDC. On the other hand, when \(n\) decreases relative to the \(p\), the LW method tends to massively under-disperse the first PC. While the other methods might overdisperse the first PC, LW underdisperse the first PC due to uniform shrinkage. This highlights why regularized PDC methods are generally preferred over LW in high-dimensional settings where \(n < p\) offers more reliable and accurate PCA results. Tables 8 and 9 highlight the challenges of PCA in high-dimensional settings where \(n < p\), particularly concerning the magnitude of PCs. Mathematically, PCs are vectors, possessing both magnitude and direction. While Tables 8 and 9 address the magnitude, Table 5 examines how closely the sample PCs align with the population parameters direction-wise.\\

\pmb{Table 5:} Average CSE of the first PC
\begin{table}[ht]
\centering
\resizebox{\columnwidth}{!}{%
\begin{tabular}{@{}|c|c|cccc
>{\columncolor[HTML]{FFFC9E}}c 
>{\columncolor[HTML]{9AFF99}}c cc|@{}}
\toprule
\cellcolor[HTML]{D9D9D9} &
  \cellcolor[HTML]{D9D9D9} &
  \multicolumn{8}{c|}{\cellcolor[HTML]{D9D9D9}\textbf{Average CSE of first PC}} \\ \cmidrule(l){3-10} 
\multirow{-2}{*}{\cellcolor[HTML]{D9D9D9}\textbf{p}} &
  \multirow{-2}{*}{\cellcolor[HTML]{D9D9D9}\textbf{n}} &
  \multicolumn{1}{c|}{\cellcolor[HTML]{D9D9D9}\textbf{POP}} &
  \multicolumn{1}{c|}{\cellcolor[HTML]{D9D9D9}\textbf{MLE}} &
  \multicolumn{1}{c|}{\cellcolor[HTML]{D9D9D9}\textbf{LW}} &
  \multicolumn{1}{c|}{\cellcolor[HTML]{D9D9D9}\textbf{PDC}} &
  \multicolumn{1}{c|}{\cellcolor[HTML]{D9D9D9}\textbf{SPDC}} &
  \multicolumn{1}{c|}{\cellcolor[HTML]{D9D9D9}\textbf{LSPDC}} &
  \multicolumn{1}{c|}{\cellcolor[HTML]{D9D9D9}\textbf{MAXPDC}} &
  \cellcolor[HTML]{D9D9D9}\textbf{RPDC} \\ \midrule
20 &
  3 &
  \multicolumn{1}{c|}{0.00000} &
  \multicolumn{1}{c|}{0.21556} &
  \multicolumn{1}{c|}{0.31244} &
  \multicolumn{1}{c|}{0.21556} &
  \multicolumn{1}{c|}{\cellcolor[HTML]{FFFC9E}0.18558} &
  \multicolumn{1}{c|}{\cellcolor[HTML]{CBCEFB}0.19727} &
  \multicolumn{1}{c|}{0.21556} &
  \cellcolor[HTML]{9AFF99}0.19138 \\ \midrule
20 &
  4 &
  \multicolumn{1}{c|}{0.00000} &
  \multicolumn{1}{c|}{0.13307} &
  \multicolumn{1}{c|}{0.18955} &
  \multicolumn{1}{c|}{0.13307} &
  \multicolumn{1}{c|}{\cellcolor[HTML]{CBCEFB}0.11066} &
  \multicolumn{1}{c|}{\cellcolor[HTML]{FFFC9E}0.10841} &
  \multicolumn{1}{c|}{0.13307} &
  \cellcolor[HTML]{9AFF99}0.10967 \\ \midrule
20 &
  5 &
  \multicolumn{1}{c|}{0.00000} &
  \multicolumn{1}{c|}{0.08324} &
  \multicolumn{1}{c|}{0.10860} &
  \multicolumn{1}{c|}{0.08324} &
  \multicolumn{1}{c|}{\cellcolor[HTML]{9AFF99}0.05233} &
  \multicolumn{1}{c|}{\cellcolor[HTML]{CBCEFB}0.05470} &
  \multicolumn{1}{c|}{0.08324} &
  \cellcolor[HTML]{FFFC9E}0.05159 \\ \midrule
20 &
  6 &
  \multicolumn{1}{c|}{0.00000} &
  \multicolumn{1}{c|}{0.06510} &
  \multicolumn{1}{c|}{0.06591} &
  \multicolumn{1}{c|}{0.06510} &
  \multicolumn{1}{c|}{\cellcolor[HTML]{FFFC9E}0.03795} &
  \multicolumn{1}{c|}{\cellcolor[HTML]{CBCEFB}0.03939} &
  \multicolumn{1}{c|}{0.06510} &
  \cellcolor[HTML]{9AFF99}0.03871 \\ \midrule
20 &
  7 &
  \multicolumn{1}{c|}{0.00000} &
  \multicolumn{1}{c|}{0.04886} &
  \multicolumn{1}{c|}{0.04886} &
  \multicolumn{1}{c|}{0.04886} &
  \multicolumn{1}{c|}{\cellcolor[HTML]{FFFC9E}0.02026} &
  \multicolumn{1}{c|}{\cellcolor[HTML]{9AFF99}0.02101} &
  \multicolumn{1}{c|}{0.04886} &
  \cellcolor[HTML]{CBCEFB}0.02339 \\ \midrule
20 &
  8 &
  \multicolumn{1}{c|}{0.00000} &
  \multicolumn{1}{c|}{0.03329} &
  \multicolumn{1}{c|}{0.03329} &
  \multicolumn{1}{c|}{0.03329} &
  \multicolumn{1}{c|}{\cellcolor[HTML]{FFFC9E}0.01179} &
  \multicolumn{1}{c|}{\cellcolor[HTML]{9AFF99}0.01276} &
  \multicolumn{1}{c|}{0.03329} &
  \cellcolor[HTML]{CBCEFB}0.01575 \\ \midrule
20 &
  9 &
  \multicolumn{1}{c|}{0.00000} &
  \multicolumn{1}{c|}{0.02963} &
  \multicolumn{1}{c|}{0.02963} &
  \multicolumn{1}{c|}{0.02963} &
  \multicolumn{1}{c|}{\cellcolor[HTML]{FFFC9E}0.01431} &
  \multicolumn{1}{c|}{\cellcolor[HTML]{9AFF99}0.01502} &
  \multicolumn{1}{c|}{0.02963} &
  \cellcolor[HTML]{CBCEFB}0.01830 \\ \midrule
20 &
  10 &
  \multicolumn{1}{c|}{0.00000} &
  \multicolumn{1}{c|}{0.02188} &
  \multicolumn{1}{c|}{0.02188} &
  \multicolumn{1}{c|}{0.02188} &
  \multicolumn{1}{c|}{\cellcolor[HTML]{FFFC9E}0.00900} &
  \multicolumn{1}{c|}{\cellcolor[HTML]{9AFF99}0.00952} &
  \multicolumn{1}{c|}{0.02188} &
  \cellcolor[HTML]{CBCEFB}0.01344 \\ \midrule
20 &
  11 &
  \multicolumn{1}{c|}{0.00000} &
  \multicolumn{1}{c|}{0.02067} &
  \multicolumn{1}{c|}{0.02067} &
  \multicolumn{1}{c|}{0.02067} &
  \multicolumn{1}{c|}{\cellcolor[HTML]{FFFC9E}0.00862} &
  \multicolumn{1}{c|}{\cellcolor[HTML]{9AFF99}0.00899} &
  \multicolumn{1}{c|}{0.02067} &
  \cellcolor[HTML]{CBCEFB}0.01326 \\ \midrule
20 &
  12 &
  \multicolumn{1}{c|}{0.00000} &
  \multicolumn{1}{c|}{0.01885} &
  \multicolumn{1}{c|}{0.01885} &
  \multicolumn{1}{c|}{0.01885} &
  \multicolumn{1}{c|}{\cellcolor[HTML]{FFFC9E}0.00786} &
  \multicolumn{1}{c|}{\cellcolor[HTML]{9AFF99}0.00816} &
  \multicolumn{1}{c|}{0.01885} &
  \cellcolor[HTML]{CBCEFB}0.01232 \\ \midrule
20 &
  13 &
  \multicolumn{1}{c|}{0.00000} &
  \multicolumn{1}{c|}{0.01568} &
  \multicolumn{1}{c|}{0.01568} &
  \multicolumn{1}{c|}{0.01568} &
  \multicolumn{1}{c|}{\cellcolor[HTML]{FFFC9E}0.00681} &
  \multicolumn{1}{c|}{\cellcolor[HTML]{9AFF99}0.00712} &
  \multicolumn{1}{c|}{0.01568} &
  \cellcolor[HTML]{CBCEFB}0.01128 \\ \midrule
20 &
  14 &
  \multicolumn{1}{c|}{0.00000} &
  \multicolumn{1}{c|}{0.01479} &
  \multicolumn{1}{c|}{0.01479} &
  \multicolumn{1}{c|}{0.01479} &
  \multicolumn{1}{c|}{\cellcolor[HTML]{FFFC9E}0.00688} &
  \multicolumn{1}{c|}{\cellcolor[HTML]{9AFF99}0.00712} &
  \multicolumn{1}{c|}{0.01479} &
  \cellcolor[HTML]{CBCEFB}0.01121 \\ \midrule
20 &
  15 &
  \multicolumn{1}{c|}{0.00000} &
  \multicolumn{1}{c|}{0.01402} &
  \multicolumn{1}{c|}{0.01402} &
  \multicolumn{1}{c|}{0.01402} &
  \multicolumn{1}{c|}{\cellcolor[HTML]{FFFC9E}0.00666} &
  \multicolumn{1}{c|}{\cellcolor[HTML]{9AFF99}0.00681} &
  \multicolumn{1}{c|}{0.01402} &
  \cellcolor[HTML]{CBCEFB}0.01125 \\ \midrule
20 &
  16 &
  \multicolumn{1}{c|}{0.00000} &
  \multicolumn{1}{c|}{0.01223} &
  \multicolumn{1}{c|}{0.01223} &
  \multicolumn{1}{c|}{0.01223} &
  \multicolumn{1}{c|}{\cellcolor[HTML]{FFFC9E}0.00616} &
  \multicolumn{1}{c|}{\cellcolor[HTML]{9AFF99}0.00634} &
  \multicolumn{1}{c|}{0.01223} &
  \cellcolor[HTML]{CBCEFB}0.01060 \\ \midrule
20 &
  17 &
  \multicolumn{1}{c|}{0.00000} &
  \multicolumn{1}{c|}{0.01138} &
  \multicolumn{1}{c|}{0.01138} &
  \multicolumn{1}{c|}{0.01138} &
  \multicolumn{1}{c|}{\cellcolor[HTML]{FFFC9E}0.00604} &
  \multicolumn{1}{c|}{\cellcolor[HTML]{9AFF99}0.00619} &
  \multicolumn{1}{c|}{0.01138} &
  \cellcolor[HTML]{CBCEFB}0.01044 \\ \midrule
20 &
  18 &
  \multicolumn{1}{c|}{0.00000} &
  \multicolumn{1}{c|}{0.01068} &
  \multicolumn{1}{c|}{0.01068} &
  \multicolumn{1}{c|}{0.01068} &
  \multicolumn{1}{c|}{\cellcolor[HTML]{FFFC9E}0.00581} &
  \multicolumn{1}{c|}{\cellcolor[HTML]{9AFF99}0.00589} &
  \multicolumn{1}{c|}{0.01068} &
  \cellcolor[HTML]{CBCEFB}0.01038 \\ \midrule
20 &
  19 &
  \multicolumn{1}{c|}{0.00000} &
  \multicolumn{1}{c|}{0.00997} &
  \multicolumn{1}{c|}{0.00997} &
  \multicolumn{1}{c|}{0.00997} &
  \multicolumn{1}{c|}{\cellcolor[HTML]{FFFC9E}0.00570} &
  \multicolumn{1}{c|}{\cellcolor[HTML]{9AFF99}0.00578} &
  \multicolumn{1}{c|}{\cellcolor[HTML]{CBCEFB}0.00997} &
  0.01011 \\ \midrule
20 &
  20 &
  \multicolumn{1}{c|}{0.00000} &
  \multicolumn{1}{c|}{0.00953} &
  \multicolumn{1}{c|}{0.00953} &
  \multicolumn{1}{c|}{0.00953} &
  \multicolumn{1}{c|}{\cellcolor[HTML]{FFFC9E}0.00567} &
  \multicolumn{1}{c|}{\cellcolor[HTML]{9AFF99}0.00577} &
  \multicolumn{1}{c|}{\cellcolor[HTML]{CBCEFB}0.00953} &
  0.00994 \\ \bottomrule
\end{tabular}%
}

\begin{tabular}{l @{\hspace{15pt}} c @{\hspace{15pt}} l @{\hspace{30pt}} c @{\hspace{15pt}} l @{\hspace{30pt}} c @{\hspace{15pt}} l}
\textbf{Note:} &
\cellcolor{yellow}{\hspace{0.01cm}\vspace{0.01cm}} & Best &
\cellcolor{green}{\hspace{0.01cm}\vspace{0.01cm}} & Second Best &
\cellcolor[HTML]{CBCEFB}{\hspace{0.01cm}\vspace{0.01cm}} & Third Best \\ \hline
\end{tabular}
\end{table}

We use the CSE of the first PC to evaluate the sample PC directions. Table 5 shows that SPDC achieves the lowest CSE, indicating that its sample PC estimates are the closest to the population parameters. Following SPDC, LSPDC and RPDC also perform well, showing low CSE. Notably, all these regularized PDC methods, except for MAXPDC, demonstrate lower CSE for the first PC than others. \\

Interestingly, MAXPDC yields a similar CSE for the first PC, which is comparable to MLE. This observation highlights the effectiveness of the regularized PDC methods (excluding MAXPDC) in providing more accurate sample PC directions in \textit{n} $<$ \textit{p} high-dimensional data settings. Overall, these findings suggest that regularized PDC methods enhance the reliability of PCA by better aligning sample PCs with their actual population counterparts.\\

As vectors, we are concerned with the PCs' direction and magnitude. However, in the context of PCA, the direction of the PCs is often considered more critical because it defines the axis along which the data has the most variance. The direction of the PCs helps in understanding the structure of the data and identifying the underlying patterns. While also important, the magnitude indicates the amount of variance captured by each PC. Therefore, both aspects are essential, but the direction is usually more emphasized when interpreting PCA results.\\

Based on this understanding, SPDC achieves the lowest CSE for the first PC, indicating its superiority in preserving the direction of the PCs. This low CSE means that SPDC provides sample PCs that closely align with the actual population parameters in direction. However, when examining the percentage of variance captured by the first PC, SPDC ranks third. In terms of magnitude, MAXPDC takes the first place. Despite this, MAXPDC does not show an improvement in CSE compared to MLE, with both methods yielding similar CSE values. Consequently, MAXPDC is not considered an improvement in the magnitude and direction of the sample PCs. \\

RPDC, on the other hand, ranks second in magnitude but third in direction. When comparing both the direction and magnitude of the sample PCs, with a priority on the direction, \textbf{SPDC demonstrates the most significant improvement in PCA for \textit{n} $<$ \textit{p} high-dimensional data settings}. This makes SPDC the most effective method among those evaluated, as it provides a better balance between capturing the variance (magnitude) and aligning with the actual direction of the PCs.

\section{Experiments on Public Data}

In this study, we analyze a Phytozome dataset \cite{doePhytozome}, a resource for comparative genomics of green plants. The dataset contains gene expression data for \textit{Brachypodium distachyon}, a model grass species, under various experimental conditions. It includes gene expression levels, measured in FPKM (Fragments per Kilobase of transcript per Million mapped reads), for approx. 34,000 genes under 74 different experimental conditions. These conditions involve different plant tissues such as shoots, leaves, and stems and various experimental treatments such as light/dark cycles, cold treatments, and different nutrient supplies. The dataset is organized with genes as rows and experimental conditions as columns. A unique GeneID identifies each gene, and the conditions represent different tissue types and treatment combinations. This setup provides a comprehensive view of gene expression profiles under diverse conditions, allowing researchers to understand how genes respond to varying treatments.\\

As we experienced in simulation studies, in a scenario where the number of experimental conditions (\( n \)) is less than the number of genes (\( p \)), the covariance estimation and PCA may also be ill-conditioned in gene expression data analysis. This lead to overdispersed covariance estimation and subsequent PCs. Therefore, such a PCA model might capture more noise than meaningful biological gene patterns. This can lead to misleading interpretations, where the PCs reflect noise rather than actual gene expression profiles. Due to this concern, it is harder to generalize these misinterpreted PCs to gene expression analysis and draw actionable insights. The techniques such as DR can help address these challenges.\\

To gauge the effectiveness of our proposed methods in terms of PCA, we perform a detailed experimental study focusing on scenarios where \textit{n} $<$ \textit{p} in this gene expression data. Expressly, we set the number of dimensions (\(p\)) to 74 and varied the number of samples (\(n\)) from 5 to 15. For each \(n < p\) scenario, we simulate 100 iterations to ensure robust and statistically meaningful results.\\

This experimental study allows us to investigate the performance of our proposed methods in terms of PCA, where the number of gene expression measurements far exceeds the number of experimental conditions in high-dimensional data settings.\\

As illustrated in Table 6, the average overdispersion of the first PC was analyzed for various \(n < p\) high-dimensional settings of gene expression data. The results indicate a clear trend: as \textit{n} decreases relative to \textit{p}, the average overdispersion of the first PC increases.\\

When examining the MLE approach, it becomes evident that the overdispersion is significantly higher in these \(n < p\) settings. However, on the other hand, a widely used alternative for the MLE of covariance is the Ledoit-Wolf covariance estimation method, yielding less significant results than MLE  in these \(n < p\) high-dimensional data settings. This suggests that Ledoit-Wolf covariance estimation is unsuited for PCA in high-dimensional settings where \(n < p\).\\

Given the marginal improvement, the PDC method focuses on its regularized versions. The RPDC demonstrates the lowest average overdispersion for the first PC, making it the most effective. Apart from that, the SPDC and LSPDC methods also show a significant performance than RPDC. But it slightly lesser extent compared to the RPDC. However, both SPDC and LSPDC significantly reduce the overdispersion of first PC and provide more reliable and accurate results compared to typical estimation methods. Table 8 highlights the reliability and applicability of RPDC, SPDC, and LSPDC in enhancing PCA outcomes for \textit{n} $<$ \textit{p} high-dimensional gene expression data.\\

\newpage

\pmb{Table 6:} Average Overdispersion of the first PC | Phytozome Data
\begin{table}[ht]
\centering
\resizebox{\columnwidth}{!}{%
\begin{tabular}{@{}|c|c|ccccc
>{\columncolor[HTML]{CBCEFB}}c c
>{\columncolor[HTML]{FFFFC7}}c |@{}}
\toprule
\cellcolor[HTML]{D9D9D9} &
  \cellcolor[HTML]{D9D9D9} &
  \multicolumn{8}{c|}{\cellcolor[HTML]{D9D9D9}\textbf{Average Overdispersion of first PC}} \\ \cmidrule(l){3-10} 
\multirow{-2}{*}{\cellcolor[HTML]{D9D9D9}\textbf{p}} &
  \multirow{-2}{*}{\cellcolor[HTML]{D9D9D9}\textbf{n}} &
  \multicolumn{1}{c|}{\cellcolor[HTML]{D9D9D9}\textbf{POP}} &
  \multicolumn{1}{c|}{\cellcolor[HTML]{D9D9D9}\textbf{MLE}} &
  \multicolumn{1}{c|}{\cellcolor[HTML]{D9D9D9}\textbf{LW}} &
  \multicolumn{1}{c|}{\cellcolor[HTML]{D9D9D9}\textbf{PDC}} &
  \multicolumn{1}{c|}{\cellcolor[HTML]{D9D9D9}\textbf{SPDC}} &
  \multicolumn{1}{c|}{\cellcolor[HTML]{D9D9D9}\textbf{LSPDC}} &
  \multicolumn{1}{c|}{\cellcolor[HTML]{D9D9D9}\textbf{MAXPDC}} &
  \cellcolor[HTML]{D9D9D9}\textbf{RPDC} \\ \midrule
74 &
  5 &
  \multicolumn{1}{c|}{0.000000} &
  \multicolumn{1}{c|}{1.095071} &
  \multicolumn{1}{c|}{0.726472} &
  \multicolumn{1}{c|}{1.095071} &
  \multicolumn{1}{c|}{\cellcolor[HTML]{FFFFC7}0.407725} &
  \multicolumn{1}{c|}{\cellcolor[HTML]{CBCEFB}0.424260} &
  \multicolumn{1}{c|}{0.620553} &
  \cellcolor[HTML]{9AFF99}0.408652 \\ \midrule
74 &
  6 &
  \multicolumn{1}{c|}{0.000000} &
  \multicolumn{1}{c|}{0.862375} &
  \multicolumn{1}{c|}{1.143960} &
  \multicolumn{1}{c|}{0.862375} &
  \multicolumn{1}{c|}{\cellcolor[HTML]{FFFFC7}0.265933} &
  \multicolumn{1}{c|}{\cellcolor[HTML]{CBCEFB}0.272175} &
  \multicolumn{1}{c|}{0.392784} &
  \cellcolor[HTML]{9AFF99}0.270337 \\ \midrule
74 &
  7 &
  \multicolumn{1}{c|}{0.000000} &
  \multicolumn{1}{c|}{0.707309} &
  \multicolumn{1}{c|}{1.389852} &
  \multicolumn{1}{c|}{0.707309} &
  \multicolumn{1}{c|}{\cellcolor[HTML]{9AFF99}0.240719} &
  \multicolumn{1}{c|}{\cellcolor[HTML]{CBCEFB}0.254671} &
  \multicolumn{1}{c|}{0.350673} &
  0.237093 \\ \midrule
74 &
  8 &
  \multicolumn{1}{c|}{0.000000} &
  \multicolumn{1}{c|}{0.575631} &
  \multicolumn{1}{c|}{1.497910} &
  \multicolumn{1}{c|}{0.575631} &
  \multicolumn{1}{c|}{\cellcolor[HTML]{9AFF99}0.201135} &
  \multicolumn{1}{c|}{\cellcolor[HTML]{CBCEFB}0.207158} &
  \multicolumn{1}{c|}{0.282708} &
  0.192903 \\ \midrule
74 &
  9 &
  \multicolumn{1}{c|}{0.000000} &
  \multicolumn{1}{c|}{0.463153} &
  \multicolumn{1}{c|}{1.576263} &
  \multicolumn{1}{c|}{0.463153} &
  \multicolumn{1}{c|}{\cellcolor[HTML]{CBCEFB}0.162813} &
  \multicolumn{1}{c|}{\cellcolor[HTML]{9AFF99}0.162324} &
  \multicolumn{1}{c|}{0.211596} &
  0.157926 \\ \midrule
74 &
  10 &
  \multicolumn{1}{c|}{0.000000} &
  \multicolumn{1}{c|}{0.469671} &
  \multicolumn{1}{c|}{1.619404} &
  \multicolumn{1}{c|}{0.469671} &
  \multicolumn{1}{c|}{\cellcolor[HTML]{9AFF99}0.145126} &
  \multicolumn{1}{c|}{\cellcolor[HTML]{CBCEFB}0.150952} &
  \multicolumn{1}{c|}{0.185968} &
  0.136837 \\ \midrule
74 &
  11 &
  \multicolumn{1}{c|}{0.000000} &
  \multicolumn{1}{c|}{0.418005} &
  \multicolumn{1}{c|}{1.577441} &
  \multicolumn{1}{c|}{0.418005} &
  \multicolumn{1}{c|}{\cellcolor[HTML]{9AFF99}0.133369} &
  \multicolumn{1}{c|}{\cellcolor[HTML]{CBCEFB}0.135252} &
  \multicolumn{1}{c|}{0.173302} &
  0.125027 \\ \midrule
74 &
  12 &
  \multicolumn{1}{c|}{0.000000} &
  \multicolumn{1}{c|}{0.316212} &
  \multicolumn{1}{c|}{1.564542} &
  \multicolumn{1}{c|}{0.316212} &
  \multicolumn{1}{c|}{\cellcolor[HTML]{CBCEFB}0.108186} &
  \multicolumn{1}{c|}{\cellcolor[HTML]{9AFF99}0.106666} &
  \multicolumn{1}{c|}{0.139324} &
  0.101614 \\ \midrule
74 &
  13 &
  \multicolumn{1}{c|}{0.000000} &
  \multicolumn{1}{c|}{0.353287} &
  \multicolumn{1}{c|}{1.618222} &
  \multicolumn{1}{c|}{0.353287} &
  \multicolumn{1}{c|}{\cellcolor[HTML]{CBCEFB}0.088806} &
  \multicolumn{1}{c|}{\cellcolor[HTML]{9AFF99}0.085105} &
  \multicolumn{1}{c|}{0.115271} &
  0.084781 \\ \midrule
74 &
  14 &
  \multicolumn{1}{c|}{0.000000} &
  \multicolumn{1}{c|}{0.288695} &
  \multicolumn{1}{c|}{1.502158} &
  \multicolumn{1}{c|}{0.288695} &
  \multicolumn{1}{c|}{\cellcolor[HTML]{9AFF99}0.081009} &
  \multicolumn{1}{c|}{\cellcolor[HTML]{CBCEFB}0.083815} &
  \multicolumn{1}{c|}{0.114219} &
  0.073784 \\ \midrule
74 &
  15 &
  \multicolumn{1}{c|}{0.000000} &
  \multicolumn{1}{c|}{0.270646} &
  \multicolumn{1}{c|}{1.339662} &
  \multicolumn{1}{c|}{0.270646} &
  \multicolumn{1}{c|}{\cellcolor[HTML]{CBCEFB}0.077814} &
  \multicolumn{1}{c|}{\cellcolor[HTML]{9AFF99}0.077240} &
  \multicolumn{1}{c|}{0.124319} &
  0.072625 \\ \bottomrule
\end{tabular}%
}

\begin{tabular}{l @{\hspace{15pt}} c @{\hspace{15pt}} l @{\hspace{30pt}} c @{\hspace{15pt}} l @{\hspace{30pt}} c @{\hspace{15pt}} l}
\textbf{Note:} &
\cellcolor{yellow}{\hspace{0.01cm}\vspace{0.01cm}} & Best &
\cellcolor{green}{\hspace{0.01cm}\vspace{0.01cm}} & Second Best &
\cellcolor[HTML]{CBCEFB}{\hspace{0.01cm}\vspace{0.01cm}} & Third Best \\ \hline
\end{tabular}
\end{table}

The MAXPDC method also reduces the overdispersion of the first PC magnitude-wise. However, it does not reduce the overdispersion of the first PC as effectively as the other regularized PDC methods. Still, MAXPDC performs better than MLE and Ledoit-Wolf. Overall, RPDC is the best method for minimizing overdispersion in the first PC under \(n < p\) conditions, highlighting its potential for improving PCA in \(n < p\) high-dimensional gene expression data.\\

Table 7 presents the average percentage of explained variance for gene expression data's first PC in selected \textit{n} $<$ \textit{p} high-dimensional data settings. As discussed in Table 6, the MLE method significantly overestimates the variance of the first PC with an unnecessarily higher percentage of explained variance in high-dimensional settings where \(n < p\). When the \(n\) increases significantly more than the \(p\), the performance of MLE improves. This implies the inapplicability and inaccuracy of using MLE when \textit{n} $<$ \textit{p} high-dimensional data settings. For example, in selected \textit{n} $<$ \textit{p} high-dimensional data settings where \textit{n} = 5 and \textit{p} = 74, in the population data, the first PC should capture about 75$\%$ of the total variance, whereas MLE estimates it at about 99$\%$, indicating a significant overestimation. On the other hand, the LW method underestimates the first PC's variance. This may happen because of its uniform shrinkage approach. This shows the limitations of using LW estimation when \textit{n} $<$ \textit{p} high-dimensional data settings. It does not accurately capture the actual variance-covariance structure in \textit{n} $<$ \textit{p} HD data. These inaccuracies highlight the need for more reliable methods for PCA when dealing with \(n < p\) HD data settings.\\

\newpage
\pmb{Table 7:} Average Percentage of Explained Variance of the First PC | Phytozome
\begin{table}[ht]
\centering
\resizebox{\columnwidth}{!}{%
\begin{tabular}{@{}|c|c|ccccc
>{\columncolor[HTML]{CBCEFB}}c c
>{\columncolor[HTML]{FFFFC7}}c |@{}}
\toprule
\cellcolor[HTML]{D9D9D9} &
  \cellcolor[HTML]{D9D9D9} &
  \multicolumn{8}{c|}{\cellcolor[HTML]{D9D9D9}\textbf{Average Percentage of Explained Variance of first PC}} \\ \cmidrule(l){3-10} 
\multirow{-2}{*}{\cellcolor[HTML]{D9D9D9}\textbf{p}} &
  \multirow{-2}{*}{\cellcolor[HTML]{D9D9D9}\textbf{n}} &
  \multicolumn{1}{c|}{\cellcolor[HTML]{D9D9D9}\textbf{POP}} &
  \multicolumn{1}{c|}{\cellcolor[HTML]{D9D9D9}\textbf{MLE}} &
  \multicolumn{1}{c|}{\cellcolor[HTML]{D9D9D9}\textbf{LW}} &
  \multicolumn{1}{c|}{\cellcolor[HTML]{D9D9D9}\textbf{PDC}} &
  \multicolumn{1}{c|}{\cellcolor[HTML]{D9D9D9}\textbf{SPDC}} &
  \multicolumn{1}{c|}{\cellcolor[HTML]{D9D9D9}\textbf{LSPDC}} &
  \multicolumn{1}{c|}{\cellcolor[HTML]{D9D9D9}\textbf{MAXPDC}} &
  \cellcolor[HTML]{D9D9D9}\textbf{RPDC} \\ \midrule
74 &
  5 &
  \multicolumn{1}{c|}{74.94184} &
  \multicolumn{1}{c|}{99.27146} &
  \multicolumn{1}{c|}{55.12550} &
  \multicolumn{1}{c|}{99.27146} &
  \multicolumn{1}{c|}{\cellcolor[HTML]{FFFFC7}89.78744} &
  \multicolumn{1}{c|}{\cellcolor[HTML]{CBCEFB}90.08548} &
  \multicolumn{1}{c|}{93.25670} &
  \cellcolor[HTML]{9AFF99}89.80431 \\ \midrule
74 &
  6 &
  \multicolumn{1}{c|}{74.94184} &
  \multicolumn{1}{c|}{99.08073} &
  \multicolumn{1}{c|}{47.13994} &
  \multicolumn{1}{c|}{99.08073} &
  \multicolumn{1}{c|}{\cellcolor[HTML]{FFFFC7}88.34649} &
  \multicolumn{1}{c|}{\cellcolor[HTML]{CBCEFB}88.50289} &
  \multicolumn{1}{c|}{91.23277} &
  \cellcolor[HTML]{9AFF99}88.45704 \\ \midrule
74 &
  7 &
  \multicolumn{1}{c|}{74.94184} &
  \multicolumn{1}{c|}{98.88958} &
  \multicolumn{1}{c|}{41.37240} &
  \multicolumn{1}{c|}{98.88958} &
  \multicolumn{1}{c|}{\cellcolor[HTML]{9AFF99}88.91245} &
  \multicolumn{1}{c|}{\cellcolor[HTML]{CBCEFB}89.31161} &
  \multicolumn{1}{c|}{91.80391} &
  88.80681 \\ \midrule
74 &
  8 &
  \multicolumn{1}{c|}{74.94184} &
  \multicolumn{1}{c|}{98.27670} &
  \multicolumn{1}{c|}{37.29955} &
  \multicolumn{1}{c|}{98.27670} &
  \multicolumn{1}{c|}{\cellcolor[HTML]{9AFF99}88.73541} &
  \multicolumn{1}{c|}{\cellcolor[HTML]{CBCEFB}88.94041} &
  \multicolumn{1}{c|}{91.29501} &
  88.45020 \\ \midrule
74 &
  9 &
  \multicolumn{1}{c|}{74.94184} &
  \multicolumn{1}{c|}{97.31829} &
  \multicolumn{1}{c|}{33.66149} &
  \multicolumn{1}{c|}{97.31829} &
  \multicolumn{1}{c|}{\cellcolor[HTML]{CBCEFB}88.20886} &
  \multicolumn{1}{c|}{\cellcolor[HTML]{9AFF99}88.18895} &
  \multicolumn{1}{c|}{90.06640} &
  88.00823 \\ \midrule
74 &
  10 &
  \multicolumn{1}{c|}{74.94184} &
  \multicolumn{1}{c|}{98.84208} &
  \multicolumn{1}{c|}{30.56228} &
  \multicolumn{1}{c|}{98.84208} &
  \multicolumn{1}{c|}{\cellcolor[HTML]{9AFF99}88.22734} &
  \multicolumn{1}{c|}{\cellcolor[HTML]{CBCEFB}88.49138} &
  \multicolumn{1}{c|}{89.98104} &
  87.84238 \\ \midrule
74 &
  11 &
  \multicolumn{1}{c|}{74.94184} &
  \multicolumn{1}{c|}{98.70888} &
  \multicolumn{1}{c|}{28.77175} &
  \multicolumn{1}{c|}{98.70888} &
  \multicolumn{1}{c|}{\cellcolor[HTML]{9AFF99}88.36677} &
  \multicolumn{1}{c|}{\cellcolor[HTML]{CBCEFB}88.46122} &
  \multicolumn{1}{c|}{90.24518} &
  87.94015 \\ \midrule
74 &
  12 &
  \multicolumn{1}{c|}{74.94184} &
  \multicolumn{1}{c|}{96.62236} &
  \multicolumn{1}{c|}{26.71663} &
  \multicolumn{1}{c|}{96.62236} &
  \multicolumn{1}{c|}{\cellcolor[HTML]{CBCEFB}87.62321} &
  \multicolumn{1}{c|}{\cellcolor[HTML]{9AFF99}87.53378} &
  \multicolumn{1}{c|}{89.33294} &
  87.23200 \\ \midrule
74 &
  13 &
  \multicolumn{1}{c|}{74.94184} &
  \multicolumn{1}{c|}{98.87711} &
  \multicolumn{1}{c|}{23.71544} &
  \multicolumn{1}{c|}{98.87711} &
  \multicolumn{1}{c|}{\cellcolor[HTML]{CBCEFB}86.94224} &
  \multicolumn{1}{c|}{\cellcolor[HTML]{9AFF99}86.68952} &
  \multicolumn{1}{c|}{88.61393} &
  86.66717 \\ \midrule
74 &
  14 &
  \multicolumn{1}{c|}{74.94184} &
  \multicolumn{1}{c|}{97.46221} &
  \multicolumn{1}{c|}{23.57136} &
  \multicolumn{1}{c|}{97.46221} &
  \multicolumn{1}{c|}{\cellcolor[HTML]{9AFF99}86.87132} &
  \multicolumn{1}{c|}{\cellcolor[HTML]{CBCEFB}87.07617} &
  \multicolumn{1}{c|}{89.10710} &
  86.32691 \\ \midrule
74 &
  15 &
  \multicolumn{1}{c|}{74.94184} &
  \multicolumn{1}{c|}{97.56997} &
  \multicolumn{1}{c|}{24.59807} &
  \multicolumn{1}{c|}{97.56997} &
  \multicolumn{1}{c|}{\cellcolor[HTML]{CBCEFB}87.07507} &
  \multicolumn{1}{c|}{\cellcolor[HTML]{9AFF99}87.03026} &
  \multicolumn{1}{c|}{90.27799} &
  86.66352 \\ \bottomrule
\end{tabular}%
}

\begin{tabular}{l @{\hspace{15pt}} c @{\hspace{15pt}} l @{\hspace{30pt}} c @{\hspace{15pt}} l @{\hspace{30pt}} c @{\hspace{15pt}} l}
\textbf{Note:} &
\cellcolor{yellow}{\hspace{0.01cm}\vspace{0.01cm}} & Best &
\cellcolor{green}{\hspace{0.01cm}\vspace{0.01cm}} & Second Best &
\cellcolor[HTML]{CBCEFB}{\hspace{0.01cm}\vspace{0.01cm}} & Third Best \\ \hline
\end{tabular}
\end{table}

However, as Table 7 shows, our proposed regularized PDC methods can address these issues effectively in \textit{n} $<$ \textit{p} high-dimensional data settings. Among them, the RPDC provides the closest estimates of the population parameters compared to all other proposed techniques and traditional techniques of MLE and Ledoit-Wolf. Following the RPDC, the SPDC, and the LSPDC also perform better than those traditional methods, with slightly less improvement than the RPDC. These regularized covariance estimation methods significantly improve the finite sample estimates of PCA over the traditional approaches.\\

At last, this detailed analysis of the average percentage of explained variance for gene expression data highlights the effectiveness of regularized PDC methods in providing reliable and accurate finite sample estimates for the PCA when \textit{n} $<$ \textit{p} high-dimensional data settings. By addressing the limitations of MLE and Ledoit-Wolf, these regularized PDC methods enhance our ability to interpret and understand high-dimensional data accurately. As we discussed earlier, the direction of the sample PC estimates plays a crucial role in PCA. Table 8 presents the CSE of the first PC for each method to evaluate the goodness of fit of these directions. Similar to the distribution of the magnitude of the PCs, the direction of the PCs also deteriorates as the \(n\) decreases relative to the \(p\).\\

\newpage
\pmb{Table 8:} Average CSE of first PC | Phytozome Data
\begin{table}[ht]
\centering
\resizebox{\columnwidth}{!}{%
\begin{tabular}{@{}|c|c|cccc
>{\columncolor[HTML]{FFFFC7}}c 
>{\columncolor[HTML]{CBCEFB}}c c
>{\columncolor[HTML]{9AFF99}}c |@{}}
\toprule
\cellcolor[HTML]{D9D9D9} &
  \cellcolor[HTML]{D9D9D9} &
  \multicolumn{8}{c|}{\cellcolor[HTML]{D9D9D9}\textbf{Average CSE of first PC}} \\ \cmidrule(l){3-10} 
\multirow{-2}{*}{\cellcolor[HTML]{D9D9D9}\textbf{p}} &
  \multirow{-2}{*}{\cellcolor[HTML]{D9D9D9}\textbf{n}} &
  \multicolumn{1}{c|}{\cellcolor[HTML]{D9D9D9}\textbf{POP}} &
  \multicolumn{1}{c|}{\cellcolor[HTML]{D9D9D9}\textbf{MLE}} &
  \multicolumn{1}{c|}{\cellcolor[HTML]{D9D9D9}\textbf{LW}} &
  \multicolumn{1}{c|}{\cellcolor[HTML]{D9D9D9}\textbf{PDC}} &
  \multicolumn{1}{c|}{\cellcolor[HTML]{D9D9D9}\textbf{SPDC}} &
  \multicolumn{1}{c|}{\cellcolor[HTML]{D9D9D9}\textbf{LSPDC}} &
  \multicolumn{1}{c|}{\cellcolor[HTML]{D9D9D9}\textbf{MAXPDC}} &
  \cellcolor[HTML]{D9D9D9}\textbf{RPDC} \\ \midrule
74 &
  5 &
  \multicolumn{1}{c|}{0.000000} &
  \multicolumn{1}{c|}{0.247691} &
  \multicolumn{1}{c|}{0.247691} &
  \multicolumn{1}{c|}{0.247691} &
  \multicolumn{1}{c|}{\cellcolor[HTML]{FFFFC7}0.076225} &
  \multicolumn{1}{c|}{\cellcolor[HTML]{CBCEFB}0.079755} &
  \multicolumn{1}{c|}{0.247691} &
  0.076708 \\ \midrule
74 &
  6 &
  \multicolumn{1}{c|}{0.000000} &
  \multicolumn{1}{c|}{0.248536} &
  \multicolumn{1}{c|}{0.248536} &
  \multicolumn{1}{c|}{0.248536} &
  \multicolumn{1}{c|}{\cellcolor[HTML]{FFFFC7}0.073888} &
  \multicolumn{1}{c|}{\cellcolor[HTML]{CBCEFB}0.093020} &
  \multicolumn{1}{c|}{0.248536} &
  0.075063 \\ \midrule
74 &
  7 &
  \multicolumn{1}{c|}{0.000000} &
  \multicolumn{1}{c|}{0.203118} &
  \multicolumn{1}{c|}{0.203118} &
  \multicolumn{1}{c|}{0.203118} &
  \multicolumn{1}{c|}{\cellcolor[HTML]{FFFFC7}0.053610} &
  \multicolumn{1}{c|}{\cellcolor[HTML]{CBCEFB}0.073243} &
  \multicolumn{1}{c|}{0.203118} &
  0.054907 \\ \midrule
74 &
  8 &
  \multicolumn{1}{c|}{0.000000} &
  \multicolumn{1}{c|}{0.223614} &
  \multicolumn{1}{c|}{0.223614} &
  \multicolumn{1}{c|}{0.223614} &
  \multicolumn{1}{c|}{\cellcolor[HTML]{FFFFC7}0.052163} &
  \multicolumn{1}{c|}{\cellcolor[HTML]{CBCEFB}0.056960} &
  \multicolumn{1}{c|}{0.223614} &
  0.052487 \\ \midrule
74 &
  9 &
  \multicolumn{1}{c|}{0.000000} &
  \multicolumn{1}{c|}{0.198741} &
  \multicolumn{1}{c|}{0.198741} &
  \multicolumn{1}{c|}{0.198741} &
  \multicolumn{1}{c|}{\cellcolor[HTML]{FFFFC7}0.051376} &
  \multicolumn{1}{c|}{\cellcolor[HTML]{CBCEFB}0.054183} &
  \multicolumn{1}{c|}{0.198741} &
  0.053768 \\ \midrule
74 &
  10 &
  \multicolumn{1}{c|}{0.000000} &
  \multicolumn{1}{c|}{0.167903} &
  \multicolumn{1}{c|}{0.167903} &
  \multicolumn{1}{c|}{0.167903} &
  \multicolumn{1}{c|}{\cellcolor[HTML]{FFFFC7}0.050993} &
  \multicolumn{1}{c|}{\cellcolor[HTML]{CBCEFB}0.054053} &
  \multicolumn{1}{c|}{0.167903} &
  0.052861 \\ \midrule
74 &
  11 &
  \multicolumn{1}{c|}{0.000000} &
  \multicolumn{1}{c|}{0.172170} &
  \multicolumn{1}{c|}{0.172170} &
  \multicolumn{1}{c|}{0.172170} &
  \multicolumn{1}{c|}{\cellcolor[HTML]{9AFF99}0.067141} &
  \multicolumn{1}{c|}{\cellcolor[HTML]{CBCEFB}0.068879} &
  \multicolumn{1}{c|}{0.172170} &
  \cellcolor[HTML]{FFFFC7}0.053295 \\ \midrule
74 &
  12 &
  \multicolumn{1}{c|}{0.000000} &
  \multicolumn{1}{c|}{0.208089} &
  \multicolumn{1}{c|}{0.208089} &
  \multicolumn{1}{c|}{0.208089} &
  \multicolumn{1}{c|}{\cellcolor[HTML]{FFFFC7}0.051925} &
  \multicolumn{1}{c|}{\cellcolor[HTML]{9AFF99}0.054721} &
  \multicolumn{1}{c|}{0.208089} &
  \cellcolor[HTML]{CBCEFB}0.054761 \\ \midrule
74 &
  13 &
  \multicolumn{1}{c|}{0.000000} &
  \multicolumn{1}{c|}{0.188317} &
  \multicolumn{1}{c|}{0.188317} &
  \multicolumn{1}{c|}{0.188317} &
  \multicolumn{1}{c|}{\cellcolor[HTML]{FFFFC7}0.088795} &
  \multicolumn{1}{c|}{\cellcolor[HTML]{CBCEFB}0.109493} &
  \multicolumn{1}{c|}{0.188317} &
  0.091722 \\ \midrule
74 &
  14 &
  \multicolumn{1}{c|}{0.000000} &
  \multicolumn{1}{c|}{0.191797} &
  \multicolumn{1}{c|}{0.183590} &
  \multicolumn{1}{c|}{0.191797} &
  \multicolumn{1}{c|}{\cellcolor[HTML]{FFFFC7}0.069056} &
  \multicolumn{1}{c|}{\cellcolor[HTML]{9AFF99}0.071941} &
  \multicolumn{1}{c|}{0.191797} &
  \cellcolor[HTML]{CBCEFB}0.072456 \\ \midrule
74 &
  15 &
  \multicolumn{1}{c|}{0.000000} &
  \multicolumn{1}{c|}{0.195967} &
  \multicolumn{1}{c|}{0.195967} &
  \multicolumn{1}{c|}{0.195967} &
  \multicolumn{1}{c|}{\cellcolor[HTML]{FFFFC7}0.067450} &
  \multicolumn{1}{c|}{\cellcolor[HTML]{9AFF99}0.069309} &
  \multicolumn{1}{c|}{0.195967} &
  \cellcolor[HTML]{CBCEFB}0.072428 \\ \bottomrule
\end{tabular}%
}

\begin{tabular}{l @{\hspace{15pt}} c @{\hspace{15pt}} l @{\hspace{30pt}} c @{\hspace{15pt}} l @{\hspace{30pt}} c @{\hspace{15pt}} l}
\textbf{Note:} &
\cellcolor{yellow}{\hspace{0.01cm}\vspace{0.01cm}} & Best &
\cellcolor{green}{\hspace{0.01cm}\vspace{0.01cm}} & Second Best &
\cellcolor[HTML]{CBCEFB}{\hspace{0.01cm}\vspace{0.01cm}} & Third Best \\ \hline
\end{tabular}
\end{table}

A careful examination of the CSE values reveals that traditional methods yield higher CSEs than our proposed regularized versions of PDC estimation for the first PC. This indicates that the direction of the PCs estimated by existing methods could be more accurate when \(n < p\). However, as shown in Table 8, all the regularized PDC estimation methods (except MAXPDC) perform significantly better in estimating the direction of the first PC when \textit{n} $<$ \textit{p} high-dimensional data settings. Among them, the SPDC method consists of the lowest CSE for the first PC.\\

This implies the SPDC's superior capability to provide the most reliable and accurate finite sample PC estimates along with the direction of the population. When we look at more closely, the RPDC method also demonstrates substantial improvement in accuracy, significantly outperforming traditional approaches such as the MLE and Ledoit-Wolf methods. On the other hand, LSPDC method's finite sample estimates of PC are slightly less accurate than SPDC and RPDC. But, still shows considerable enhancements in the direction of PC estimation compared to traditional techniques. These results highlight the effectiveness of our regularized PDC methods, particularly SPDC, in yielding reliable and accurate finite sample PC estimations for the first PC with low CSE, affirming their superiority in \textit{n} $<$ \textit{p} high-dimensional data analysis. Figure 2 shows that, when comparing the direction and magnitude of the sample PC estimates generated by existing methods, the LW must effectively tackle the issue of overdispersion in the first PC when pitted against the MLE method. However, the MLE method itself generates sample PC estimates that are significantly overdispersed.\\

\begin{figure}[h]
\centering
\includegraphics[width=1.0\textwidth]{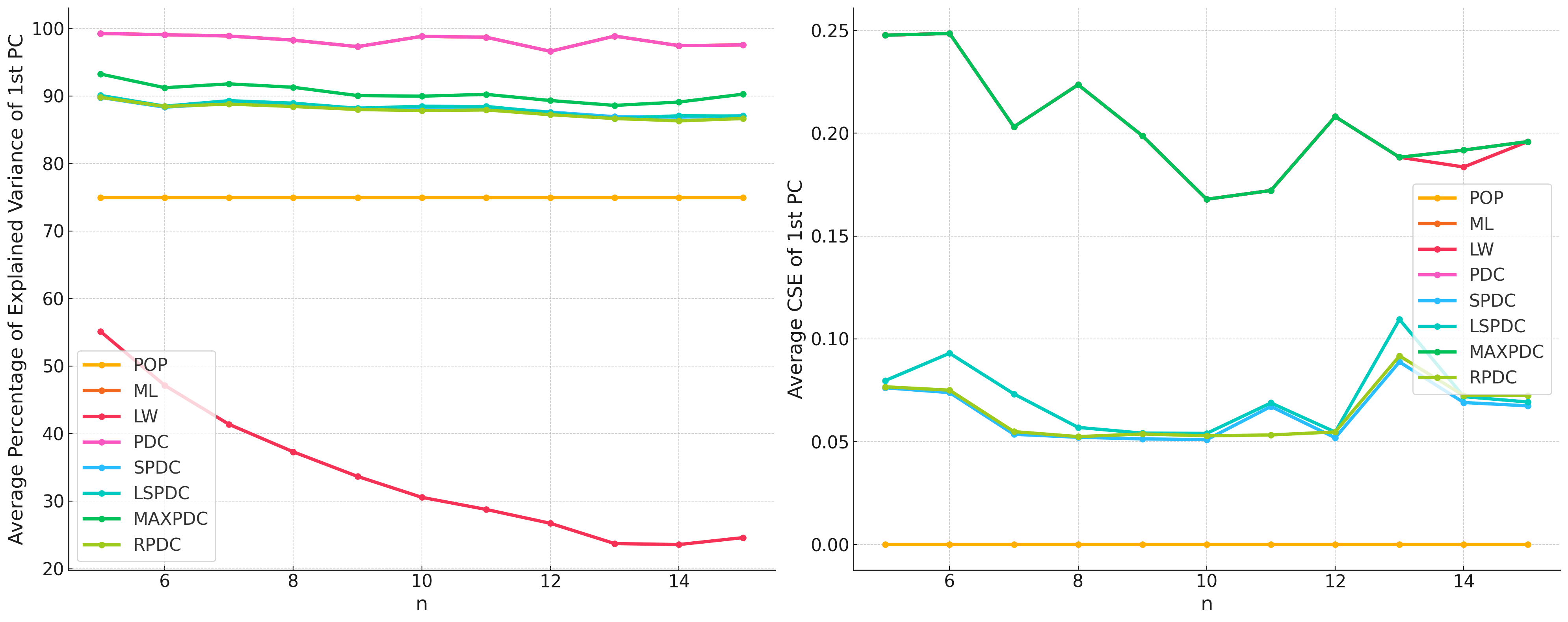}
    \caption{Method Comparisons | Phytozome Data}\label{fig52}
\end{figure}

\newpage
However, our proposed regularized versions of PDC estimations are highly effective in mitigating the overdispersion of the first PC. Among these, the SPDC method demonstrates the most substantial improvement by prioritizing the direction of the sample estimates. Following SPDC, the RPDC method also effectively addresses the overdispersion issue, albeit slightly less. Lastly, the LSPDC method provides notable improvements over MLE in managing the overdispersion of the first PC.\\

In summary, our proposed regularized PDC methods, particularly SPDC, significantly improve the accuracy of PCA in \textit{n} $<$ \textit{p} high-dimensional settings by reducing the overdispersion of the first PC, thereby offering a more reliable and robust alternative to traditional methods such as MLE and Ledoit-Wolf covariance estimation.

\section{Results and Discussion}

This study looks at how well PCA works with different covariance estimation methods in high-dimensional data settings where \textit{n} $<$ \textit{p}. In this context, traditional covariance estimation techniques often struggle due to the issue of overdispersion of the sample estimates. This leads to the need to explore more robust and reliable methods of covariance estimation for \textit{n} $<$ \textit{p} high-dimensional data settings. To minimize the overdispersion of PCA when \textit{n} $<$ \textit{p} high-dimensional data settings, we suggest a new covariance estimation method called PDC along with its four regularized versions: SPDC, LSPDC, MAXPDC, and RPDC. These methods are benchmarked against traditional techniques such as MLE covariance estimation and Ledoit-Wolf covariance estimation, widely regarded as reliable standards in practical applications. In this study, we evaluate how these methods perform in terms of improving the reliability and accuracy of PCA results when \textit{n} $<$ \textit{p} high-dimensional data settings.\\

Ledoit-Wolf covariance estimation is renowned for enhancing the robustness and stability of the explained variance in PCA. It achieves this by applying a shrinkage technique that combines the sample covariance matrix with a structured estimator, typically a multiple of the identity matrix. This approach overcomes the risk of overfitting and reduces sampling variability. However, it leads to underdispersion of first \textit{n} - 1 PCs and overdisoersion of last \textit{p}-\textit{n}+1 PCs due to its uniform shrinkage of sample eigenvalues. Moreover, the Ledoit-Wolf method primarily targets stabilizing eigenvalues of the sample covariance matrix rather than adjusting the direction of the eigenvectors or PCs. This is because the shrinkage technique focuses on regularizing the magnitude of the covariance estimates, thereby improving the overall robustness of the explained variance without fundamentally changing the angles of the PCs.\\

On the other hand, the proposed PDC estimation method introduces a novel approach that consistently and marginally reduces the overdispersion of the first PC when \textit{n} $<$ \textit{p} high-dimensional data settings. However, these estimated components are slightly closer to the MLE method’s sample estimates. Therefore, this improvement is marginally small (on the order of $10^{-15}$), providing a more balanced variance distribution across all PCs. This improvement is significant in high-dimensional settings where the first PC often dominates the explained variance. By achieving a more balanced variance distribution, PDC enhances PCA performance in terms of both the magnitude and direction of the PCs. The too small but persistent improvements offered by PDC make it a valuable addition to the toolkit of covariance estimation methods for PCA when \textit{n} $<$ \textit{p} high-dimensional data settings.\\

All the regularized versions of PDC, SPDC, LSPDC, MAXPDC, and RPDC, demonstrate significant improvements over traditional covariance estimation methods. Among these, SPDC stands out for its ability to effectively regularize the magnitude and direction of the sample PCs compared to MLE. This method of covariance estimations significantly reduces the overdispersion of the first few PCs, as SPDC redistributes the excess variance from the first \textit{n} - 1 PCs more evenly among the subsequent PCs. The enhanced regularization provided by SPDC leads to a more balanced spread of variance, thereby improving the overall PCA performance. Following SPDC, RPDC, LSPDC, and MAXPDC also show substantial improvements, each contributing unique strengths to the regularization process.

\bibliographystyle{plain}
\bibliography{References.bib}

\end{document}